\documentclass[twocolumn]{aastex63}
\usepackage{mathrsfs}
\usepackage{color}
\usepackage{comment}
\usepackage{url}
\usepackage{graphicx} % Including figure files
\usepackage{amsmath}  % Advanced maths commands
\usepackage{amssymb}  % Extra maths symbols
\usepackage{bm}   % Bold maths symbols, including upright Greek
\usepackage{array}
\usepackage{booktabs}
\usepackage{longtable}
\usepackage{supertabular}
\usepackage{threeparttable}

\received{March 2, 2021}
\submitjournal{AJ}

\shorttitle{Thermophysical modeling of Themis Family}
\shortauthors{Jiang \& Ji}

\graphicspath{{./}{figures/}}

\begin{document}

\title{Thermophysical Modeling of 20 Themis Family Asteroids with WISE/NEOWISE Observations}

\correspondingauthor{Jianghui Ji}
\email{jijh@pmo.ac.cn, jianghx@pmo.ac.cn}

\author{Haoxuan Jiang}
\affil{CAS Key Laboratory of Planetary Sciences, Purple Mountain Observatory, Chinese Academy of Sciences, Nanjing 210023, China\\}
\author{Jianghui Ji}
\affil{CAS Key Laboratory of Planetary Sciences, Purple Mountain Observatory, Chinese Academy of Sciences, Nanjing 210023, China\\}
\affil{CAS Center for Excellence in Comparative Planetology, Hefei 230026, China\\}

\begin{abstract}
Themis family is one of the largest and oldest asteroid populations in the main-belt. Water-ice may widely exist on the parent body (24) Themis. In this work, we employ the Advanced Thermophysical Model as well as mid-infrared measurements from NASA's Wide-Field Infrared Survey Explorer to explore thermal parameters of 20 Themis family members. Here we show that the average thermal inertia and geometric albedo are ~$39.5\pm26.0 ~\rm J m^{-2} s^{-1/2} K^{-1}$ and $0.067\pm0.018$, respectively. The family members have a relatively moderate roughness fraction on their surfaces. We find that the relatively low albedos of Themis members are consistent with the typical values of B-type and C-type asteroids. As aforementioned, Themis family bears a very low thermal inertia, which indicates a fine and mature regolith on their surfaces. The resemblance of thermal inertia and geometric albedo of Themis members may reveal their close connection in origin and evolution. In addition, we present the compared results of thermal parameters for several prominent families.

\end{abstract}


\section{Introduction} \label{sec:intro}
The Themis family is located in the outer region of the asteroid belt at a mean distance of 3.13 AU from the Sun, and is one of the oldest asteroid family, having been predicted to be formed $\sim 2.5\pm1.0~\rm Gyr$~ago \citep{2013A&A...551A.117B} by a collisional event from its parent body (24) Themis. In Nesvorn\'y Hierarchical Clustering Method (HCM) Asteroid Family Catalog, Themis family is one of the largest asteroid populations, which includes more than 4700 family members~\citep{2012PDSS..189.....N} with proper orbital elements $3.08\leq a_{\rm p} \leq 3.24$ AU, $0.09 \leq e_{\rm p} \leq 0.22$ and $i_{\rm p} \le 3^\circ$, where $a_{\rm p}$, $e_{\rm p}$, and $i_{\rm p}$ are proper semi-major axis, eccentricity and inclination, respectively. This population is known as one of the most statistically reliable asteroid family in the main belt~\citep{2016Icar..269....1F}.

Most of the Themis family members are recognized as C-type asteroids \citep{2005Icar..174...54M}. \citet{2011Icar..213..538Z} presented near-infrared spectra ($0.8-2.4~\rm \mu m$) of 7 Themis family asteroids, and found that the spectra of carbonaceous chondrite meteorites can fit the general spectral shapes and trends of their Themis family members. Based on NASA 3.0-m Infrared Telescope Facility (IRTF), \citet{2010Natur.464.1322R} and \citet{2010Natur.464.1320C} reported the spectroscopic detection of water-ice and organic material on the family's parent body (24) Themis, indicating the widespread presence of water-ice in asteroidal surfaces and interiors of Themis.  IRTF's spectra showed the presence of a constant depth in $3.1-\rm \mu m$ absorption due to the existence of water-ice, as well as absorptions between 3.3 and 3.6 $\rm \mu m$ that are closely matched by organic compounds, implying that the ice and organic are widespread and may be evenly distributed over the surface of (24) Themis \citep{2010Natur.464.1320C}. If (24) Themis had experienced cometary activities or impact event, the surface ice may be replenished by a sub-surface reservoir \citep{2010Natur.464.1320C}. Moreover, \citet{2016Icar..269....1F} presented the outcomes of visible and near-infrared spectroscopic survey of 22 Themis family members, and found these asteroids have diverse spectral behaviors including blue/neutral and moderately red spectra, which 4 of them showed absorption bands centered at $0.68-0.73 ~\rm \mu m$, indicating the presence of aqueous alteration. Besides, \citet{2016A&A...586A..15M} analysed near-infrared spectral properties of 15 Themis family members, which are found to be consistent with that of chondritic porous interplanetary dust particles, and ultra-fine grained materials are found to be the dominant constituents, thereby inferring a parent body accreted from a mixture of ice and anhydrous silicates.

Furthermore, water-ice was also discovered on two main belt comets (MBCs), 133P/Elst-Pizarro and 176P/LINEAR, which connected with Themis family from viewpoints of dynamical evolution and spectral reflectance \citep{2012A&A...537A..73L,2006Sci...312..561H}.   Dynamical analysis showed that MBCs are more likely to have formed in situ in the main belt, rather than originate from the outer solar system \citep{2002Icar..159..358F}. In particular, if they are the fragments of a collisional family, the activities of MBCs are driven by water-ice sublimation, implying that a plenty of asteroids from Themis family may have water-ice under the surface. As a matter of fact, after water-ice on (24) Themis was detected by \citet{2010Natur.464.1322R} and \citet{2010Natur.464.1320C}, there are also similar water-ice detections for other MBAs, along with 4 Themis family members \citep{2012Icar..219..641T,2015Icar..254..150H}. Moreover, the visible spectra of 133P/Elst-Pizarro and 176P/LINEAR have resemblance to those of three Themistians, indicating the two MBCs may be the member of Themis family \citep{2011A&A...532A..65L}.   With the data of Spitzer Space Telescope, \citet{2009ApJ...694L.111H} determined the geometric R-band albedos and effective diameters of two MBCs, $p_{\rm R,133P}=0.05\pm0.02$, $D_{\rm 133P}=3.8\pm0.3~\rm km$ and $p_{\rm R,176P}=0.06\pm0.02$, $D_{\rm 176P}=4.0\pm0.2~\rm km$. Recently, \citet{2020AJ....159...66Y} derived the geometric albedo and effective diameter of 133P/Elst-Pizarro to be $0.074\pm0.013$ and $3.9_{-0.3}^{+0.4}~\rm km$ , respectively, and evaluated the thermal inertia of 133P/Elst-Pizarro to be 25 ~$\rm J m^{-2} s^{-1/2} K^{-1}$. The geometric albedos of two MBCs correspond roughly to the typical values of Themistians. Therefore, the Themis family members may contain crucial clue to catastrophic event and interior characteristics of their parent body.

The cometary activities of Themistians may be partly involved in the surface temperature, which can be determined by certain thermophysical models. \citet{2013ApJ...770....7M} applied the Near Earth Asteroid Thermal Model (NEATM) \citep{1998Icar..131..291H} to investigate thermal characteristics for a wide variety of asteroid families. However, the NEATM model can simply obtain the albedo and diameter of the asteroid, whereas the heat conduction and temperature variation procedure is dominated by thermal inertia. Thus, we need to employ a more sophisticated thermal model to understand thermal features of Themis family. In this work, we investigate 20 Themistians from the perspective of thermal physics. Thus we aim to derive their geometric albedos, thermal inertia, effective diameters, roughness fraction, and obtain their distribution characteristics. The results may be considered to assess whether the asteroids are "interlopers", thereby revealing the homogeneity/heterogeneity of the family members. More recently, by using the data of Subaru and Herschel telescopes, \citet{2020ApJ...898L..45O} showed that the thermal inertia and geometric albedo of (24) Themis to be $\Gamma_{\rm Themis}=20_{-10}^{+25}$ ~$\rm J m^{-2} s^{-1/2} K^{-1}$, $p_{\rm v,Themis}=0.07\pm0.01$ , respectively, with a diameter $D_{\rm Themis}=192_{-7}^{+10}~\rm km$.  Moreover, the asteroid families are formed from the impact events in a wide variety of ages and heliocentric distances, thereby making spacial environment of the families diversified, which may in turn have induced an evolution of thermal process. The physical nature of the family members are also determined by the materials of parent body and the impactor. Therefore, by comparing the variations in thermal parameters of individual family, one may infer the collisional scenario of the families and the characteristics of the parent body. In addition, thermal inertia distribution among individual asteroid family may be a crucial evidence for the existence of asteroidal differentiation \citep{2013Icar..226..419M}. As iron meteorites have higher thermal conductivity than ordinary and carbonaceous chondrites \citep{2010Icar..208..449O}, a metal iron-rich regolith is expected to have larger thermal inertia, thus thermal inertia can help us distinguish iron-rich or iron-poor asteroids \citep{2015aste.book..107D}. Table~\ref{orbit_themis} lists the target asteroids in this work, where includes the orbital and physical parameters as well as the spectral type. From Table~\ref{orbit_themis}, we note that except (1633) Chimay and (1687) Glarona, the remaining bodies are  B-type or C-type asteroids, and  (2592) Hunan is a slow rotator with a rotation period of approximately 50 hours when compared to others.

This paper is structured as follows. In Section 2 we introduce the Advanced Thermophysical Model (ATPM), which can be used to calculate the theoretical flux of the target asteroids. The radiometric results for 20  Themistians under study and their analysis are presented in Section 3. In Section 4, we show the distribution of thermal parameters and compare those with other asteroid families. Section 5 summarizes the results.

\section{Advanced Thermophysical Model and WISE observations}
As mentioned in \citet{2015aste.book..107D}, asteroid thermophysical modeling aims to calculate the temperature of asteroids' surface by using specific thermophysical model, then the theoretical flux emitted by the asteroid can be obtained from Planck function. The temperature is determined by several thermal process such as the absorption of sunlight, multiple scattering and reflected thermal emission, as well as the heat conduction into the subsurface. By comparing the theoretical flux and observational flux, we can constrain thermal parameters such as thermal inertia, geometric albedo, effective diameter, as well as roughness fraction. Thermal inertia plays a vital role in dominating the thermal conduction procedure on the surface of asteroid, which can be written as
\begin{equation}
  \Gamma=\sqrt{\kappa\rho C}
\label{definegamma}
\end{equation}
where $\kappa,\rho$ and $C$ represent the thermal conductivity, bulk density, and specific heat capacity of the asteroid, respectively. The thermal inertia is an intrinsic parameter that depends on the characteristics of surface component. Since $\kappa$ is a function of temperature, thermal inertia is associated with the asteroid's surface temperature. The presence of thermal inertia leads to surface temperature peaks at afternoon, as well as non-zero temperature at the nightside~\citep{2015aste.book..107D}. In addition, thermal inertia plays a significant part in the Yarkovsky and Yarkovsky-O'Keefe-Radzievskii-Paddack (YORP) effects, which can make the semi-major axis of asteroids drift and alter their spin rate ~\citep{2007Icar..190..236D}.

Here we adopt the Advanced Thermophysical Model (ATPM) \citep{2011MNRAS.415.2042R,2017MNRAS.472.2388Y} to evaluate the temperature distribution and thermal emission from the asteroids. In ATPM, an asteroid is treated as a polyhedron composed of a number of triangular facets, and a hemispherical crater is also adopted to represent the rough surface. All  shape models that we employ can be retrieved from the Database of Asteroid Models from Inversion Techniques (DAMIT)\footnote{https://astro.troja.mff.cuni.cz/projects/asteroids3D/web.php} and are determined by the light curve inversion method developed by \citet{2002Icar..159..369K}. Moreover, the thermal observations we utilize can be acquired from the Wide-Field Infrared Survey Explorer (WISE) database~\citep{wirght2010}.

In order to obtain the distribution of temperature $T$ on the asteroid's surface, we need to solve the 1D heat conduction equation on each shape facet~\citep{2015aste.book..107D}:
\begin{equation}
    \frac{\partial T}{\partial t}=\frac{\kappa}{\rho C}\frac{\partial^2 T}{\partial x^2},
  \label{heatchonduct}
\end{equation}
where $t$ is time, $x$ is the depth below the asteroid surface, $\kappa,\rho$ and $C$ are given as in Eq.~\ref{definegamma}. Considering the upper and lower boundary conditions:

\begin{equation}
    \begin{aligned}
(1-A_{\rm B})([1-S(t)]\psi(t)F_{\rm sun}+F_{\rm scat})+(1-A_{\rm th})F_{\rm rad}   \\
+ ~\kappa\Big(\frac{dT}{dx}\Big)_{x=0}=\varepsilon\sigma T^4_{x=0},  
\end{aligned}
  \label{upperboundary}
\end{equation}
\begin{equation}
\frac{\partial T}{\partial x}\Big|_{x\to\infty}=0,
\label{lowerboundary}
\end{equation}
where $A_{B}$ is the bond albedo, $S(t)$ indicates whether the facet is shadowed at time $t$, $\varepsilon$ is the thermal emissivity, $\psi$ is the cosine value of the solar altitude, $A_{\rm th}$ is the albedo at specific thermal infrared wavelength. $F_{\rm sun}$, $F_{\rm scat}$ and $F_{\rm rad}$ represents the incident sunlight, multi-scattered and re-emitted thermal flux from other facets, respectively. Eq.\ref{lowerboundary} indicates that when it is deep enough below the asteroid's surface, the temperature variation tends to be zero. In addition, we adopt the method given in \citet{jianghx2019} to remove the portion of reflected sunlight in short wavelengths (such as W1 band of WISE observations).

Furthermore, we download thermal-infrared data from 3 source tables of the WISE archive (http://irsa.ipac.caltech.edu/applications/wise/), WISE All-Sky Single Exposure (L1b), WISE 3-Band Cryo Single Exposure (L1b) Source Table and NEOWISE-R Single Exposure (L1b) Source Table. As well-known, WISE surveyed the sky with 4 wavebands centered at 3.4, 4.6, 12.0 and 22.0 $\mu \rm m$, noted as W1, W2, W3 and W4, respectively, while NEOWISE only observes at W1 and W2 bands when the solid hydrogen cryosat run out. We employ the Moving Object Catalog Search with a search cone radius of $1''$. We adopt similar criteria described in \citet{2012ApJ...744..197G} to screen the dataset that the artifact identification flag cc\_flag other than 0 and p (which indicates the source is unaffected by known artifacts), the photometric quality flag ph\_qual other than A, B, and C (which indicates that the source is likely to have been a valid detection that have signal-to-noise ratio $>2$), solar system object association flag sso\_flg other than 1 (which means the source is associated with the predicted position of a known solar system object) are rejected. Moreover, as mentioned in \citet{2020AJ....159..264J}, for main-belt asteroids, the surface temperature is low enough that the observed data in W1 band contains a significant part ($\sim 90\%$) of reflected sunlight, indicating that the thermal part only covers $\sim 10\%$ of the total observed flux. Thus, we do not use W1 data in our fitting process. Although the W2 observation also contains a significant part of reflected sunlight, the contribution is less than $50\%$. Therefore, to cover a wider range of solar phase angles and wavelengths to improve the reliability of fitting results, here we adopt W2 data, as described in \citet{jianghx2019} and \citet{2020AJ....159..264J}, to account for the reflected sunlight during fitting procedure.

\makeatletter\def\@captype{table}\makeatother
\begin{table*}
  %\centering
  \setlength{\arraycolsep}{0.2pt}
  %\small
  \caption{Orbital and physical parameters of the investigated Themis members (epoch JD=2459000.5, MPC)}
  \label{orbit_themis}
  \begin{tabular}{lcccccccc} % four columns, alignment for each
    \hline
    \hline
    Asteroid & $a$ (au) & $e$ & $i$ ($^\circ$) & $P_{\rm orb}$ (yr) & $P_{\rm rot}$ (hr) & Pole ($^\circ$) & H & Spec.type\\
    \hline
    \specialrule{0em}{1.5pt}{1.5pt}
    (62) Erato & 3.1286 & 0.1677 & 2.2366 & 5.53 & 9.2182 & (87,22) & 8.78 & ${\rm B^t/C^b}$ \\
    \specialrule{0em}{1.5pt}{1.5pt}
    (171) Ophelia & 3.1301 & 0.1320 & 2.5468 & 5.54 & 6.6645 & (144,29) & 8.60 & ${\rm C^t/C^b}$\\
    \specialrule{0em}{1.5pt}{1.5pt}
    (222) Lucia & 3.1430 & 0.1312 & 2.1490 & 5.57 & 7.8367 & (293,49) & 9.63 & ${\rm B^t}$ \\
    \specialrule{0em}{1.5pt}{1.5pt}
    (468) Lina & 3.1332 & 0.1974 & 0.4369 & 5.55 & 15.4784 & (74,68) & 9.76 & ${\rm C^t}$ \\
    \specialrule{0em}{1.5pt}{1.5pt}
    (526) Jena & 3.1208 & 0.1335 & 2.1737 & 5.51 & 11.8765 & (194,54) & 10.17 & ${\rm B^t/C^s}$ \\
    \specialrule{0em}{1.5pt}{1.5pt}
    (767) Bondia & 3.1220 & 0.1822 & 2.4118 & 5.52 & 8.3376 & (106,15) & 10.20 & ${\rm C^b}$ \\
    \specialrule{0em}{1.5pt}{1.5pt}
    (936) Kunigunde & 3.1323 & 0.1762 & 2.3660 & 5.54 & 8.8265 & (234,50) & 10.45 & ${\rm B^{st}/B^{sb}}$ \\
    \specialrule{0em}{1.5pt}{1.5pt}
    (996) Hilaritas & 3.0901 & 0.1398 & 0.6589 & 5.43 & 10.0515 & (281,-57) & 11.19 & ${\rm B^t}$ \\
    \specialrule{0em}{1.5pt}{1.5pt}
    (1082) Pirola & 3.1241 & 0.1813 & 1.8525 & 5.52 & 15.8540 & (123,-42) & 10.48 & ${\rm C^t}$\\
    \specialrule{0em}{1.5pt}{1.5pt}
    (1576) Fabiola & 3.1429 & 0.1671 & 0.9542 & 5.57 & 6.8891 & (229,75) & 11.15 & ${\rm B^t}$ \\%BU
    \specialrule{0em}{1.5pt}{1.5pt}
    (1633) Chimay & 3.1929 & 0.1238 & 2.6764 & 5.71 & 6.5906 & (116,81) & 10.75 & $\rm S^*$ \\
    \specialrule{0em}{1.5pt}{1.5pt}
    (1687) Glarona & 3.1688 & 0.1722 & 2.6358 & 5.64 & 6.4960 & (132,76) & 10.70 & $\rm S^*$ \\
    \specialrule{0em}{1.5pt}{1.5pt}
    (1691) Oort & 3.1636 & 0.1757 & 1.0860 & 5.63 & 10.2684 & (223,58) & 10.98 & ${\rm C^t}$ \\%CU
    \specialrule{0em}{1.5pt}{1.5pt}
    (2528) Mohler & 3.1472 & 0.1708 & 0.5119 & 5.58 & 6.4918 & (56,-64) & 12.28 & $\rm C^*$ \\
    \specialrule{0em}{1.5pt}{1.5pt}
    (2592) Hunan & 3.1205 & 0.1232 & 1.3369 & 5.51 & 49.9871 & (184,-73) & 12.22 & $\rm C^*$ \\
    \specialrule{0em}{1.5pt}{1.5pt}
    (2659) Millis & 3.1317 & 0.1029 & 1.3214 & 5.54 & 6.1246 & (109,-49) & 11.77 & $\rm B^b/C^s$ \\
    \specialrule{0em}{1.5pt}{1.5pt}
    (2673) Lossignol & 3.2055 & 0.1511 & 2.2773 & 5.74 & 4.9379 & (274,44) & 12.55 & $\rm C^*$ \\
    \specialrule{0em}{1.5pt}{1.5pt}
    (2708) Burns & 3.0821 & 0.1777 & 2.7828 & 5.41 & 5.3236 & (183,-59) & 12.17 & $\rm B^b$ \\
    \specialrule{0em}{1.5pt}{1.5pt}
    (2718) Handley & 3.1201 & 0.1553 & 1.4921 & 5.51 & 13.0980 & (261,-53) & 11.89 & $\rm C^s$ \\
    \specialrule{0em}{1.5pt}{1.5pt}
    (2803) Vilho & 3.1410 & 0.1790 & 1.3300 & 5.57 & 10.3728 & (285,-67) & 12.01 & $\rm C^*$ \\
    \specialrule{0em}{1.5pt}{1.5pt}
    \hline
  \end{tabular}
  \begin{tablenotes}
    \footnotesize
     \item [] Notes:  $a$: semi-major axis, $e$: eccentricity, $i$: orbital inclination, $P_{\rm orb}$: orbital period, H: absolute magnitude are from the Minor Planet Center (MPC)). $P_{\rm rot}$: rotation period, Pole: orientation are obtained \citet{2011A&A...530A.134H,2013A&A...551A..67H,2016A&A...586A.108H}, \citet{2019A&A...625A.139M}, and \citet{2016A&A...587A..48D,2018A&A...617A..57D,2019A&A...631A...2D}. Spectral types with superscript t: Tholen taxonomic classification \citep{1989aste.conf.1139T},  b: Bus-Demeo and Bus-Binzel taxonomic classification \citep{2002Icar..158..146B,2009Icar..202..160D}, s: SDSS taxonomic classification \citep{2010A&A...510A..43C}, st and sb: Tholen-like and Bus-like in $\rm S^3OS^2$ taxonomic classification \citep{2004Icar..172..179L} and *: spectral types in Asteroid Lightcurve Database (LCDB). 
    \end{tablenotes}
\end{table*}

\makeatletter\def\@captype{table}\makeatother
\begin{table*}
  \centering
  \small
  \caption{Derived thermal parameters of Themis family asteroids based on ATPM and WISE/NEOWISE observations}
  \label{resultsall_themis}
  \begin{tabular}{lccccccc} % four columns, alignment for each
\hline
\hline
Asteroid & $\Gamma$ ($\rm J m^{-2} s^{-1/2} K^{-1}$)& $p_{\rm v}$ &$p_{\rm v}^*$& $D_{\rm eff}$ (km)&$D_{\rm eff}^*$ (km)&$ f_{\rm r}$&$\chi^2_{\rm min}$\\
\hline
\specialrule{0em}{1.5pt}{1.5pt}															
(62) Erato 	&	 $55_{-12}^{+12}$	&	$0.0890_{-0.0110}^{+0.0070}$ 	&	 $0.0910\pm0.0020$	&	$81.064_{-3.011}^{+5.528}$ 	&	$78.620\pm0.900$	&	$0.5_{-0.2}^{+0.0}$ 	&	1.865	\\
\specialrule{0em}{1.5pt}{1.5pt}															
(171) Ophelia	&	$30_{-11}^{+15}$	&	$0.0595_{-0.0060}^{+0.0060}$	&	$0.0773\pm0.0198$	&	$103.816_{-4.869}^{+5.667}$	&	$104.103\pm1.389$	&	$0.5_{-0.1}^{+0.0}$	&	3.522	\\
\specialrule{0em}{1.5pt}{1.5pt}															
(222) Lucia 	&	 $70_{-20}^{+22}$	&	$0.0670_{-0.0085}^{+0.0110}$	&	$0.1233\pm0.0177$	&	$61.729_{-4.518}^{+4.332}$	&	$56.520\pm0.832$	&	$0.4_{-0.1}^{+0.1}$	&	2.385	\\
\specialrule{0em}{1.5pt}{1.5pt}															
(468) Lina 	&	 $5_{-5}^{+23}$ 	&	 $0.0520_{-0.0035}^{+0.0025}$	&	$0.0488\pm0.0342$ 	&	$66.915_{-1.552}^{+2.372}$ 	&	 $59.673\pm18.220$	&	 $0.1_{-0.1}^{+0.2}$	&	2.211	\\
\specialrule{0em}{1.5pt}{1.5pt}															
(526) Jena 	&	$10_{-10}^{+16}$ 	&	$0.0530_{-0.0070}^{+0.0065}$	&	$0.0580\pm0.0177$	&	$55.120_{-3.098}^{+4.046}$	&	$51.032\pm0.742$	&	$0.0_{-0.0}^{+0.1}$	&	7.395	\\
\specialrule{0em}{1.5pt}{1.5pt}															
(767) bondia 	&	 $65_{-17}^{+30}$ 	&	 $0.0575_{-0.0095}^{+0.0095}$ 	&	 $0.0900\pm0.0200$ 	&	 $50.546_{-3.720}^{+4.776}$ 	&	 $45.300\pm4.500$	&	 $0.2_{-0.1}^{+0.1}$ 	&	2.854	\\
\specialrule{0em}{1.5pt}{1.5pt}															
(936) Kunigunde 	&	 $65_{-25}^{+20}$ 	&	 $0.0749_{-0.0237}^{+0.0500}$	&	 $0.0650\pm0.1400$ 	&	$40.391_{-9.113}^{+8.462}$ 	&	 $42.230\pm1.040$ 	&	 $0.5_{-0.0}^{+0.0}$ 	&	3.871	\\
\specialrule{0em}{1.5pt}{1.5pt}															
(996) Hilaritas 	&	 $50_{-26}^{+16}$ 	&	 $0.0770_{-0.0090}^{+0.0120}$ 	&	 $0.0824\pm0.0180$ 	&	 $27.560_{-1.925}^{+1.767}$ 	&	 $30.902\pm0.417$ 	&	 $0.5_{-0.3}^{+0.0}$ 	&	8.555	\\
\specialrule{0em}{1.5pt}{1.5pt}															
(1082) Pirola 	&	 $45_{-12}^{+14}$ 	&	 $0.0725_{-0.0055}^{+0.0075}$ 	&	 $0.0867\pm0.0105$ 	&	 $41.054_{-1.972}^{+1.651}$ 	&	 $37.363\pm1.036$ 	&	 $0.5_{-0.1}^{+0.0}$ 	&	5.621	\\
\specialrule{0em}{1.5pt}{1.5pt}															
(1576) Fabiola 	&	 $10_{-10}^{+27}$ 	&	 $0.0830_{-0.0130}^{+0.0135}$ 	&	 $0.0746\pm0.0139$ 	&	 $27.796_{-2.018}^{+2.471}$ 	&	 $30.150\pm0.400$ 	&	 $0.0_{-0.0}^{+0.3}$ 	&	3.172	\\
\specialrule{0em}{1.5pt}{1.5pt}															
(1633) Chimay 	&	 $30_{-13}^{+18}$ 	&	 $0.0405_{-0.0060}^{+0.0065}$ 	&	 $0.0785\pm0.0135$ 	&	 $47.849_{-3.431}^{+3.993}$ 	&	 $37.732\pm0.426$ 	&	 $0.1_{-0.1}^{+0.2}$ 	&	5.479	\\
\specialrule{0em}{1.5pt}{1.5pt}															
(1687) Glarona 	&	 $15_{-10}^{+23}$ 	&	 $0.1155_{-0.0090}^{+0.0065}$ 	&	 $0.0795\pm0.0130$ 	&	 $34.852_{-0.941}^{+1.442}$ 	&	 $42.007\pm0.515$ 	&	 $0.0_{-0.0}^{+0.5}$ 	&	6.475	\\
\specialrule{0em}{1.5pt}{1.5pt}															
(1691) Oort 	&	 $10_{-10}^{+19}$ 	&	 $ 0.0635_{-0.0085}^{+0.0085}$ 	&	 $0.0672\pm0.0150$ 	&	 $34.844_{-2.121}^{+2.595}$ 	&	 $33.163\pm0.534$ 	&	 $0.3_{-0.1}^{+0.2}$ 	&	7.921	\\
\specialrule{0em}{1.5pt}{1.5pt}															
(2528) Mohler 	&	 $35_{-28}^{+15}$ 	&	 $0.0764_{-0.0090}^{+0.0063}$ 	&	 $0.0567\pm0.0045$ 	&	 $16.826_{-0.654}^{+1.088}$ 	&	 $19.443\pm0.121$ 	&	 $0.5_{-0.5}^{+0.0}$ 	&	10.084	\\
\specialrule{0em}{1.5pt}{1.5pt}															
(2592) Hunan 	&	 $40_{-40}^{+46}$ 	&	 $0.0636_{-0.0060}^{+0.0087}$ 	&	 $0.0724\pm0.0035$ 	&	 $18.958_{-1.177}^{+0.963}$ 	&	 $18.533\pm0.107$ 	&	 $0.3_{-0.3}^{+0.2}$ 	&	6.823	\\
\specialrule{0em}{1.5pt}{1.5pt}															
(2659) Millis 	&	 $35_{-16}^{+19}$ 	&	 $0.0645_{-0.0072}^{+0.0125}$ 	&	 $0.0498\pm0.0028$ 	&	 $27.463_{-2.328}^{+1.674}$ 	&	 $27.878\pm0.337$ 	&	 $0.5_{-0.3}^{+0.0}$ 	&	6.231	\\
\specialrule{0em}{1.5pt}{1.5pt}															
(2673) Lossignol 	&	 $15_{-15}^{+29}$ 	&	 $0.0860_{-0.0147}^{+0.0117}$ 	&	 $0.0773\pm0.0140$ 	&	 $14.005_{-0.865}^{+1.376}$ 	&	 $15.119\pm0.160$ 	&	 $0.2_{-0.2}^{+0.3}$ 	&	5.967	\\
\specialrule{0em}{1.5pt}{1.5pt}															
(2708) Burns 	&	 $65_{-48}^{+25}$ 	&	 $0.0570_{-0.0097}^{+0.0110}$ 	&	 $0.0836\pm0.0151$ 	&	 $20.492_{-1.731}^{+2.003}$ 	&	 $20.085\pm0.110$	&	 $0.5_{-0.5}^{+0.0}$ 	&	8.218	\\
\specialrule{0em}{1.5pt}{1.5pt}															
(2718) Handley 	&	 $30_{-30}^{+16}$ 	&	 $0.0519_{-0.0073}^{+0.0038}$ 	&	 $0.0550\pm0.0054$ 	&	 $24.431_{-0.849}^{+1.924}$ 	&	 $25.929\pm0.234$ 	&	 $0.5_{-0.4}^{+0.0}$ 	&	8.069	\\
\specialrule{0em}{1.5pt}{1.5pt}															
(2803) Vilho 	&	 $110_{-29}^{+12}$ 	&	 $0.0360_{-0.0010}^{+0.0071}$ 	&	 $0.0732\pm0.0104$ 	&	 $27.757_{-2.389}^{+0.394}$ 	&	 $21.441\pm0.301$ 	&	 $0.5_{-0.2}^{+0.0}$ 	&	5.218	\\
\specialrule{0em}{1.5pt}{1.5pt}
\hline
\end{tabular}
\begin{tablenotes}
    \footnotesize
    \item[] Notes: $p_{\rm v}^*$ and $D_{\rm eff}^*$ are the previous results of geometric albedo and effective diameters from \citet{2011ApJ...741...90M,2012ApJ...759L...8M,2016AJ....152...63N,
        2016A&A...591A..14A,2017AJ....154..168M}.
\end{tablenotes}
\end{table*}

\section{Results of Themis family's thermal parameters}
For convenience, we transform Eq.\ref{heatchonduct} and the boundary condition Eq.~\ref{upperboundary} into a dimensionless form (see \citet{1989aste.conf..128L} for details), which can be expressed as a function of $\Gamma$. As MBAs usually have small thermal inertia, we set the search range of $\Gamma$ to be $0\sim200$~$\rm J m^{-2} s^{-1/2} K^{-1}$ at equally spaced steps of 5 $\rm J m^{-2} s^{-1/2} K^{-1}$ during our fitting process. When Eq.\ref{heatchonduct} is solved, the radiance at observational wavelength $\lambda$ can be calculated by the Planck function,
\begin{equation}
  B(\lambda,T)=\frac{2hc^2}{\lambda^5}\frac{1}{e^{\frac{hc}{\lambda k_{\rm b} T}} - 1},
\label{planckfun}
\end{equation}
where $h$ is the Planck constat, $c$ is the speed of light and $k_{\rm b}$ is the Boltzmann constant. Thus the total theoretical thermal flux observed by the telescope can be written as:
\begin{equation}
  F_{\lambda,\rm th} = (1-f_{\rm r}) \sum_{i=1}^{N}\pi \varepsilon f_{i} S_i B(\lambda,T_i) 
  + f_{\rm r} \sum_{\rm i=1}^{N}\sum_{\rm j = 1}^{M} \pi \varepsilon B(\lambda,T_{ij}) S_{ij} f_{ij},
\label{theore_flux}
\end{equation}
where $f_i$ ($f_{ij}$), $S_i$ ($S_{ij}$) and $T_i$ ($T_{j}$) are the view factor, surface area and temperature of facet $i$ (subfacet $ij$ of the roughness facet) in the shape model, $f_{\rm r}$ is the roughness fraction that denotes the fractional coverage of hemispherical craters. We adopt the reduced $\chi^2$ introduced in \citet{numericalrecipe} to evaluate the fitting degree,
\begin{equation}
  \chi^2_{\rm r} = \frac{1}{n-3}\sum_{i=1}^{n}\left[\frac{{\rm FCF} * F_{\rm model}(\lambda_{i})-F_{obs}(\lambda_{i})}{\sigma_{\lambda,i}}\right]^2,
\label{chi2fitting}
\end{equation}
where $n$ is the number of observed data points, $\rm FCF$ is the flux correction factor that is  related to roughness fraction $f_{\rm r}$ and bond albedo $A_{\rm B}$ ($A_{\rm B}=p_{\rm v}\times qph$, $qph$ is the phase integral), which is introduced in \citet{2011MNRAS.418.1246W}. It should be emphasized that $F_{\rm model}$ in Eq.\ref{chi2fitting} includes the thermal part $F_{\rm th}$ in Eq.\ref{theore_flux} and the reflected sunlight contribution (see \citet{jianghx2019} and \citet{2020AJ....159..264J} for details). Besides, the effective diameter $D_{\rm eff}$ and geometric albedo $p_{\rm v}$ are correlated by \citep{fowler1992}
\begin{equation}
  D_{\rm eff} = \frac{1329\times10^{-H/5}}{\sqrt{p_{\rm v}}},
  \label{eqdpv}
\end{equation}
and can be treated as a single free parameter. Thus, we totally have three free parameters $\Gamma$, $p_{\rm v}$ ($D_{\rm eff}$) and $f_{\rm r}$ in Eq.\ref{chi2fitting}. In the following section, we report the fitting results of thermal parameters of 20 Themis family asteroids.

\begin{figure*}
  \centering
  \includegraphics[scale=0.58]{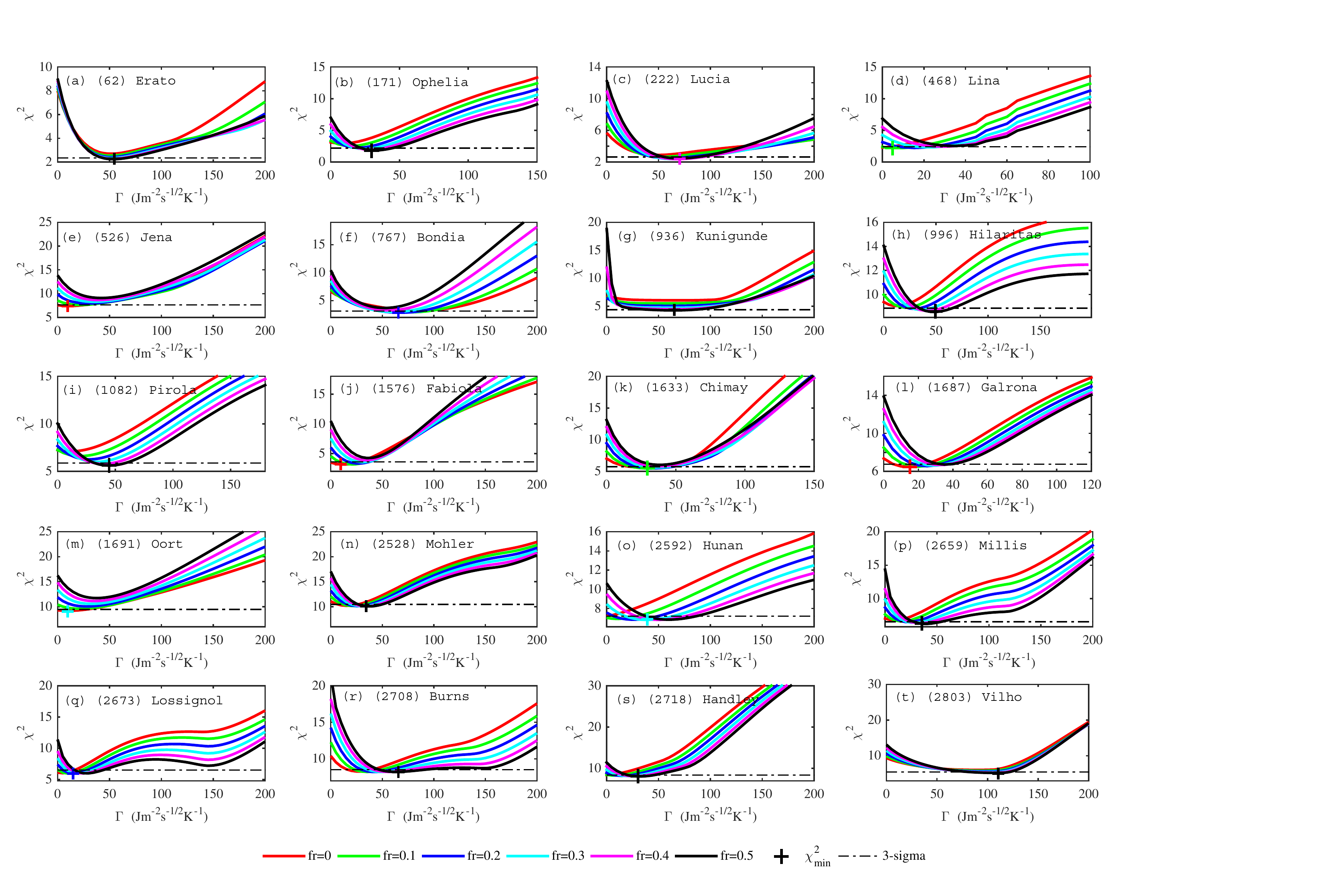}
    \caption{The $\Gamma -\chi^2$ profiles of the 20 Themis family members. The solid lines in different colors represents the roughness fraction, and the 3-$\sigma$ range of $\Gamma$ is constrained by the horizontal dashed line. The minimum value of $\chi^2$ are marked with '+' , which has the same color as the corresponding $f_{\rm r}$.}
    \label{chi2gaall}
\end{figure*}

\subsection{(62) Erato}
Asteroid (62) Erato is a relatively large member in the Themis family with an absolute magnitude of 8.78. \citet{2017AJ....154..168M} presented the diameter and albedo of this asteroid to be $92.197\pm27.20~\rm km$ and $0.1016\pm0.1021$, while \citet{2016AJ....152...63N} obtained the diameter of $80.09\pm 24.95~\rm km$, and geometric albedo of $0.07\pm0.11$.  Here we adopt 136 W2 observations from NEOWISE to evaluate the thermal parameters of (62) Erato in our fitting. As illustrated in  Figure~\ref{chi2gaall}(a), we can obtain the best-fitting value of thermal inertia to be $55_{-12}^{+12}$~$\rm J m^{-2} s^{-1/2} K^{-1}$ and roughness fraction $0.5_{-0.2}^{+0.0}$ ($3~\sigma$ error bars) with a minimum $\chi^2$ of 1.865, and give a visible geometric albedo of $0.0890_{-0.0110}^{+0.0070}$ and an effective diameter of $81.064_{-3.011}^{+5.528}~\rm km$. The results of $p_{\rm v}$ and $D_{\rm eff}$ are close to those of \citet{2016AJ....152...63N}.

\begin{figure*}
  \centering
  \includegraphics[scale=0.40]{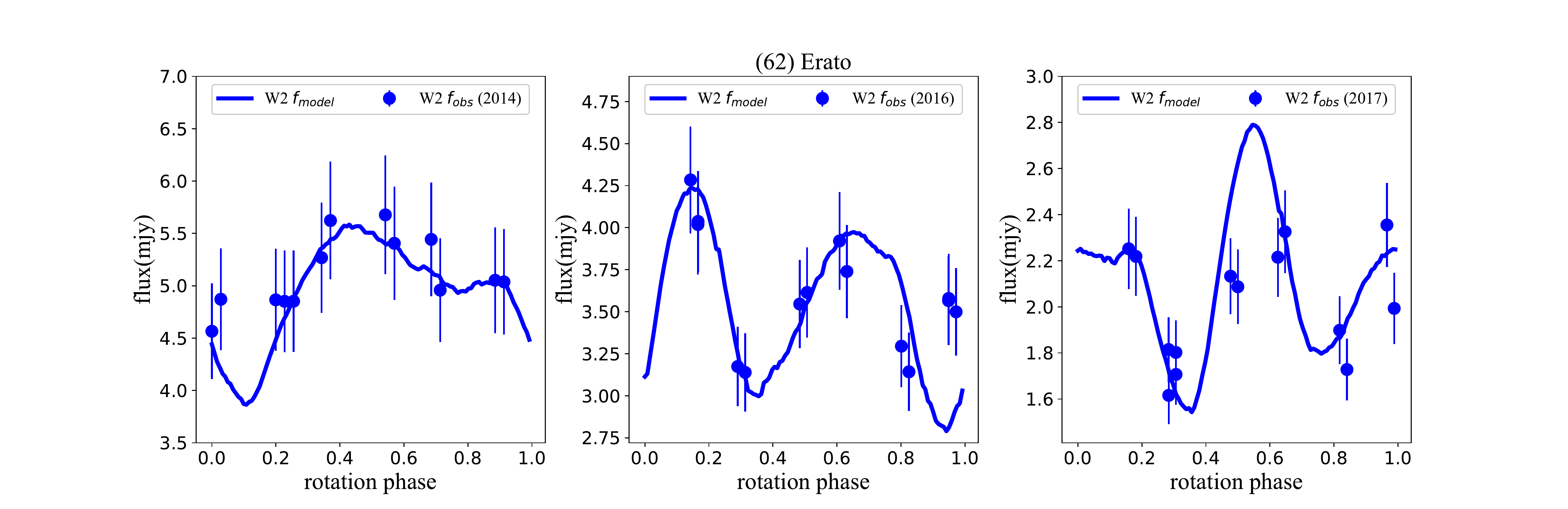}
    \caption{$\rm W2$ thermal light curves of (62) Erato.}
    \label{thli62}
\end{figure*}

To verify the best-fitting parameters for (62) Erato, here we employ the method of \citet{2017MNRAS.472.2388Y} and \citet{jianghx2019} to exhibit the theoretical thermal light curves of (62) Erato as compared with WISE/NEOWISE observations.  Since there are no W3 and W4 observations for this asteroid, only W2 thermal light curves of three years are shown in Figure~\ref{thli62}. The shape of these curves are different because we use various reference epoch of the zero rotation phase.

\subsection{(171) Ophelia}
\begin{figure*}
  \centering
  \includegraphics[scale=0.40]{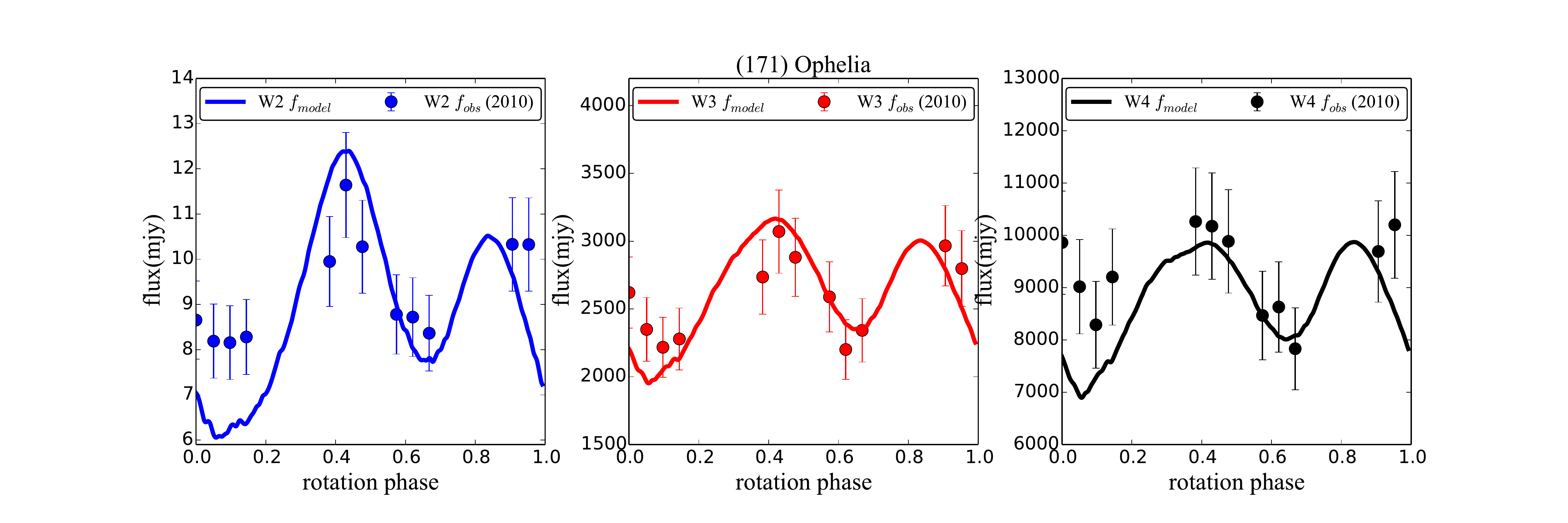}
    \caption{W2, W3 and W4 thermal light curves of (171) Ophelia.}
    \label{thli171}
\end{figure*}

 Using 181 WISE/NEOWISE observations (157 in W2, 12 in W3 and 12 in W4), we compute the geometric albedo and effective diameter of Ophelia to be $0.0595_{-0.0060}^{+0.0060}$ and $103.816_{-4.869}^{+6.667}~\rm km$, respectively, which agrees with those of $0.0773\pm0.0198$ and $104.103\pm1.389~\rm km$ from \citet{2011ApJ...741...90M}. In addition, we find that thermal inertia of $\Gamma=30_{-11}^{+15}$ ~$\rm J m^{-2} s^{-1/2} K^{-1}$ and a roughness fraction of $0.5_{-0.0}^{+0.1}$ ~(Figure~\ref{chi2gaall}(b)), which perform a good fitting with the observed data. As shown in Figure~\ref{thli171}, we plot the theoretical thermal light curves with respect to the observations at W2, W3 and W4 in 2010, here we use different colors (hereafter, blue for W2, red for W3 and black for W4, respectively) to represent different wavelengths, where we can observe that the ATPM results can reasonably fit most mid-infrared data points, but seem to underestimate the observations at low rotation phases, and the $\chi^2_{\rm min}$ is 3.522.

\subsection{(222) Lucia}
\begin{figure*}
  \centering
  \includegraphics[scale=0.40]{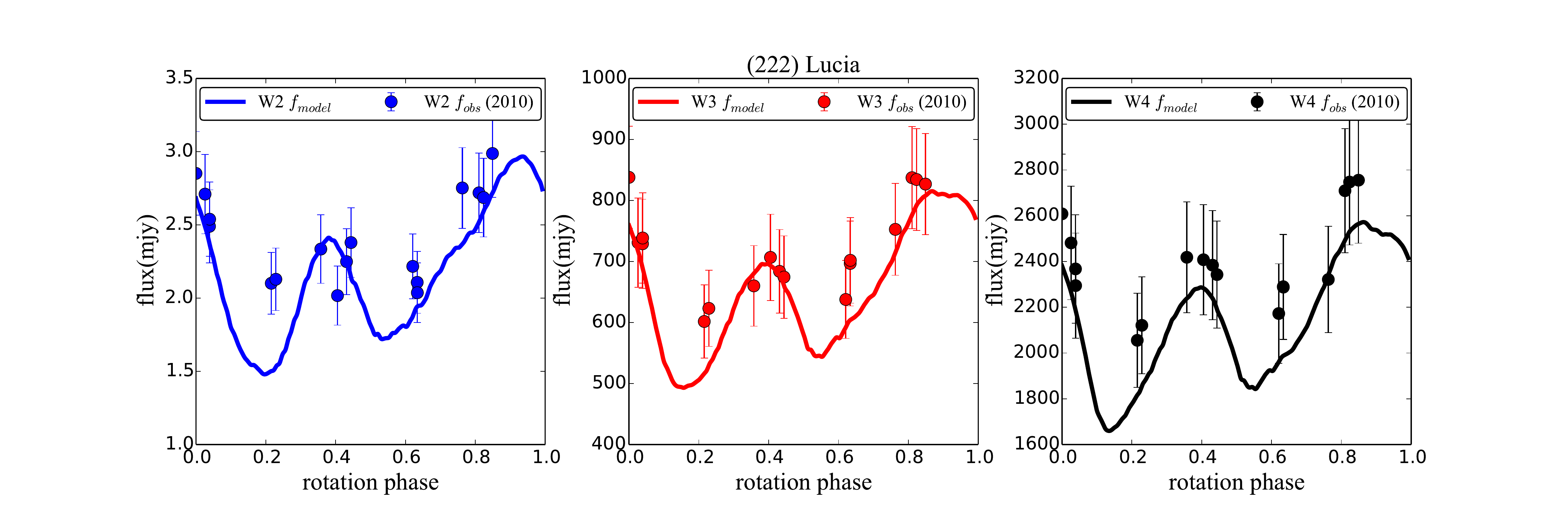}
    \caption{W2, W3 and W4 thermal light curves of (222) Lucia.}
    \label{thli222}
\end{figure*}

 Lucia was well observed by WISE/NEOWISE, \citet{2011ApJ...741...90M} derived the $p_{\rm v}$ and $D_{\rm eff}$ to be $0.1233\pm0.0177$ and $56.52\pm0.832~\rm km$, here we entirely use 193 data points (159 in W2, 17 in W3 and 17 in W4) combined with the ATPM model to derive its thermal parameters.  As can be seen from Figure~\ref{chi2gaall}(c), the best-fitting thermal inertia and roughness are $70_{-20}^{+22}$~$\rm J m^{-2} s^{-1/2} K^{-1}$ and $0.4_{-0.1}^{+0.1}$, while $p_{\rm v}$ and $D_{\rm eff}$ are constrained to be $0.0670_{-0.0085}^{+0.0110}$ and $61.729_{-4.518}^{+4.332}~\rm km$, respectively. Our results of $p_{\rm v}$ and $D_{\rm eff}$ are different from those of \citet{2011ApJ...741...90M}, which may be caused by the usage of different thermal model. Thermal light curves of W2, W3 and W4 are exhibited in Figure~\ref{thli222}, in which the modeled fluxes are slightly smaller than the observations, and the minimum value of $\chi^2$ is 2.385.

\subsection{(468) Lina}
\begin{figure*}
  \centering
  \includegraphics[scale=0.40]{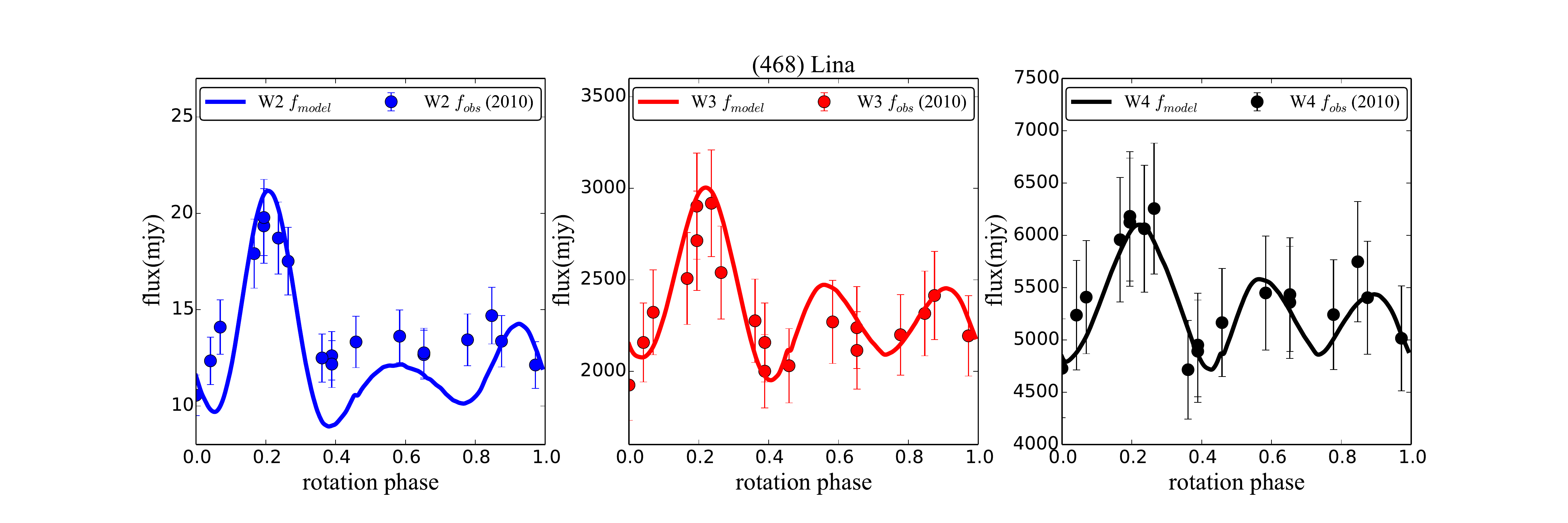}
    \caption{W2, W3 and W4 thermal light curves of (468) Lina.}
    \label{thli468}
\end{figure*}

 \citet{2019A&A...625A.139M} provided the thermal inertia of (468) Lina $\Gamma=20$~$\rm J m^{-2} s^{-1/2} K^{-1}$ with the data from WISE (W4 band), IRAS and AKARI, while \citet{2017AJ....154..168M} used the NEATM model and WISE observation to obtain the $p_{\rm v}$ and $D_{\rm eff}$ to be $0.0488\pm0.0342$ and $59.676\pm18.22~\rm km$. Here we adopt 206 WISE/NEOWISE observations (164 in W2, 21 in W3 and 21 in W4) in our fitting, and we obtain a low thermal inertia of $5_{-5}^{+23}$~$\rm J m^{-2} s^{-1/2} K^{-1}$, and a small roughness of ${0.1_{-0.1}^{+0.2}}$ (Figure~\ref{chi2gaall}(d)). Our derived  geometric albedo $p_{\rm v}=0.0520_{-0.0035}^{+0.0025}$ is slightly larger than that of \citet{2017AJ....154..168M}, thereby giving rise to a diameter of $66.915_{-1.552}^{+2.372}~\rm km$. The 3-bands thermal light curves are plotted in Figure~\ref{thli468}. Our ATPM fluxes can reasonably match the observations in W3 and W4, and the $\chi^2_{\rm min}$ is constrained to be 2.211.

\subsection{(526) Jena}
\begin{figure*}
  \centering
  \includegraphics[scale=0.40]{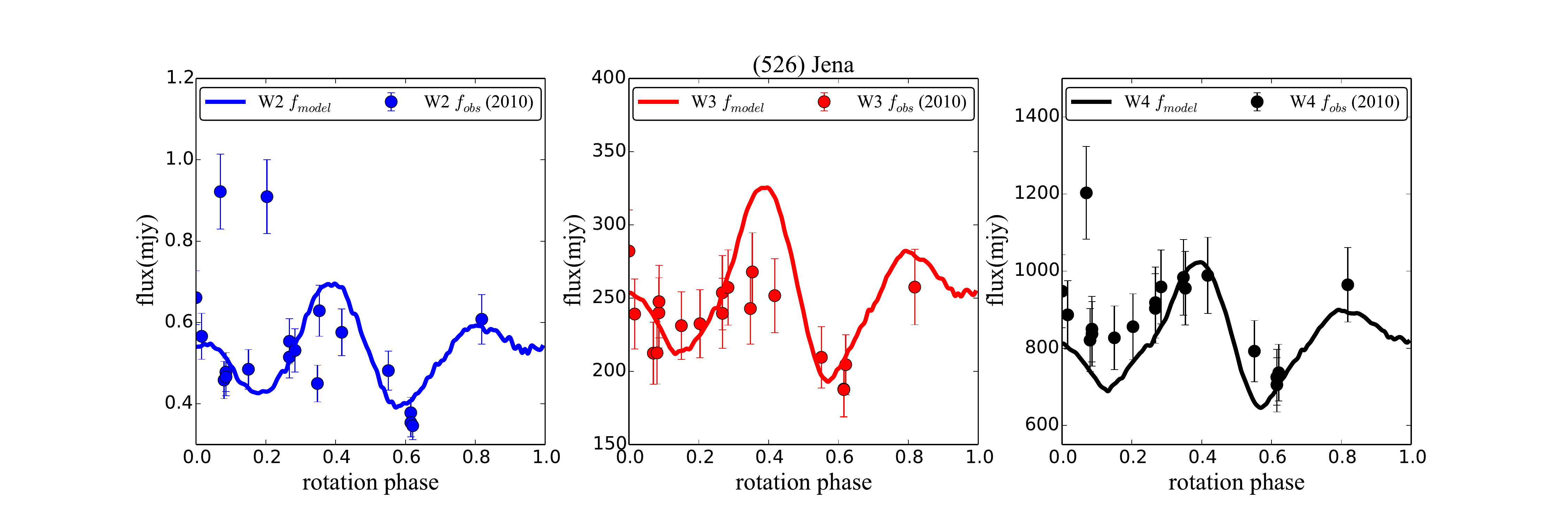}
    \caption{W2, W3 and W4 thermal light curves of (526) Jena.}
    \label{thli526}
\end{figure*}

\citet{2011ApJ...741...90M} predicted the albedo and diameter of $0.0580\pm0.0177$ and $51.032\pm0.742~\rm km$ by using the NEATM model, which are very close to those of \citet{2012A&A...537A..73L}. In this study, we fit the ATPM fluxes with 190 WISE/NEOWISE observations (150 in W2, 20 in W3 and 20 in W4). The best-fitting values are  $p_{\rm v}=0.0530_{-0.0070}^{+0.0065}$, $D_{\rm eff}=55.120_{-3.098}^{+4.046}~\rm km$, a low thermal inertia $\Gamma=10_{-10}^{+16}$~$\rm J m^{-2} s^{-1/2} K^{-1}$ with roughness fraction $f_{\rm r} = 0.0_{-0.0}^{+0.1}$  (see Figure~\ref{chi2gaall}(e)). Our results of geometric albedo and effective diameter of (526) Jena are consistent with those of \citet{2011ApJ...741...90M}. Figure~\ref{thli526} shows that the thermal light curves can agree well with the observations at W3, but in W2 and W4 bands, the observations are not well fitted at several rotation phases, providing $\chi^2_{\rm min}=7.395$, which indicates a relatively poor fit.

\subsection{(767) Bondia}
\begin{figure*}
  \centering
  \includegraphics[scale=0.40]{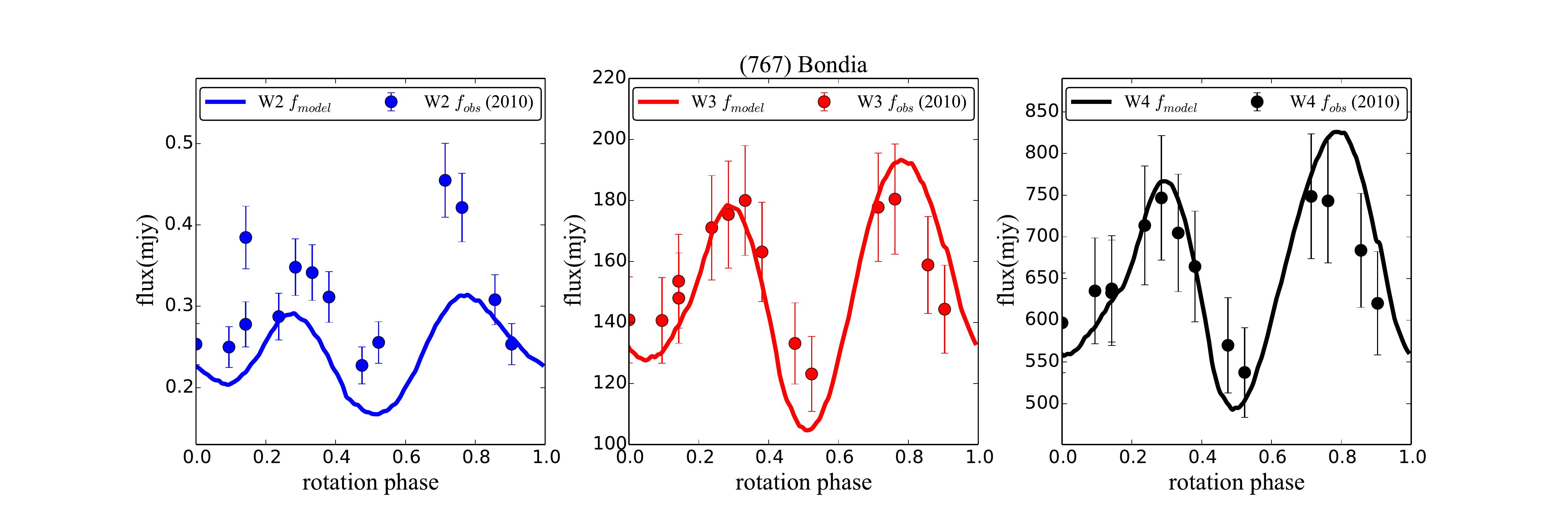}
    \caption{W2, W3 and W4 thermal light curves of (767) Bondia.}
    \label{thli767}
\end{figure*}

 The early studies individually reported $p_{\rm v}$ of $0.0956\pm0.0179$ and  $0.09\pm0.02$, along with a diameter of $43.100\pm0.730~\rm km$ and $45.3\pm4.5~\rm km$ \citep{2011ApJ...741...90M,2016A&A...591A..14A}. Here we utilize 163 WISE/NEOWISE observations (135 in W2, 14 in W3 and 14 in W4). We derive the geometric albedo $0.0575_{-0.0095}^{+0.0095}$, the diameter $50.546_{-3.720}^{+4.776}~\rm km$, the thermal roughness fraction $0.2_{-0.1}^{+0.1}$ ,and thermal inertia $65_{-17}^{+30}$~$\rm J m^{-2} s^{-1/2} K^{-1}$ (from Figure~\ref{chi2gaall}(f)). The ATPM thermal light curves supply intuitively acceptable fits with the observed data at W3 and W4 bands, but is underestimated at W2 (see Figure~\ref{thli767}), and the value of minimum $\chi^2$ is 2.854.

\subsection{(936) Kunigunde}
\begin{figure*}
  \centering
  \includegraphics[scale=0.40]{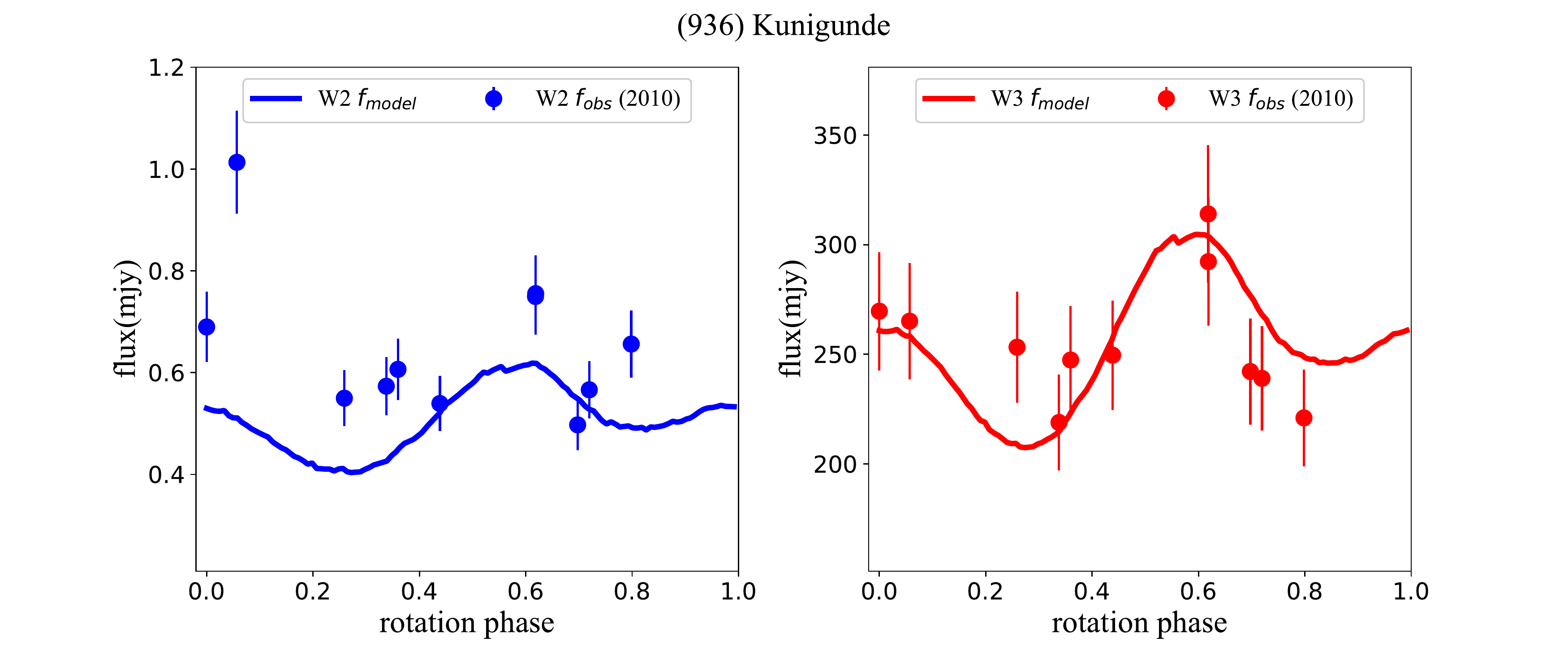}
    \caption{$\rm W2\sim W3$ thermal light curves of (936) Kunigunde.}
    \label{thli936}
\end{figure*}

Based on the measurements of space-based infrared telescopes, AKARI, IRAS (The Infrared Astronomical Satellite) and WISE, the geometric albedo of the asteroid were derived to be $0.124\pm0.007$, $0.1129\pm0.007$, $0.065\pm0.014$, respectively, with respect to each diameter of  $38.08\pm0.94$, $39.56\pm1.2$, $43.227\pm1.035~\rm km$ ~\citep{2004PDSS...12.....T,2011PASJ...63.1117U,2012ApJ...759L...8M}. In this work, 136 WISE/NEOWISE data (125 in W2 and 11 in W3) were used to calculate its thermal parameters, and we present $\Gamma = 65_{-25}^{+20}$~$\rm J m^{-2} s^{-1/2} K^{-1}$ with $f_{\rm r}=0.5_{-0.0}^{+0.0}$ (Figure~\ref{chi2gaall}(g)), $p_{\rm v}=0.0749_{-0.0237}^{+0.0500}$ and $D_{\rm eff}=40.391_{-9.113}^{+8.462}~\rm km$, respectively. Our derived best-fitting albedo and diameter are very close to those of ~\citet{2012ApJ...759L...8M}. The thermal light curves versus the data at W2, W3 are shown in Figure~\ref{thli936}, and the $\chi^2_{\rm min}$ value is calculated to be 3.871.

\subsection{(996) Hilaritas}
\begin{figure*}
  \centering
  \includegraphics[scale=0.40]{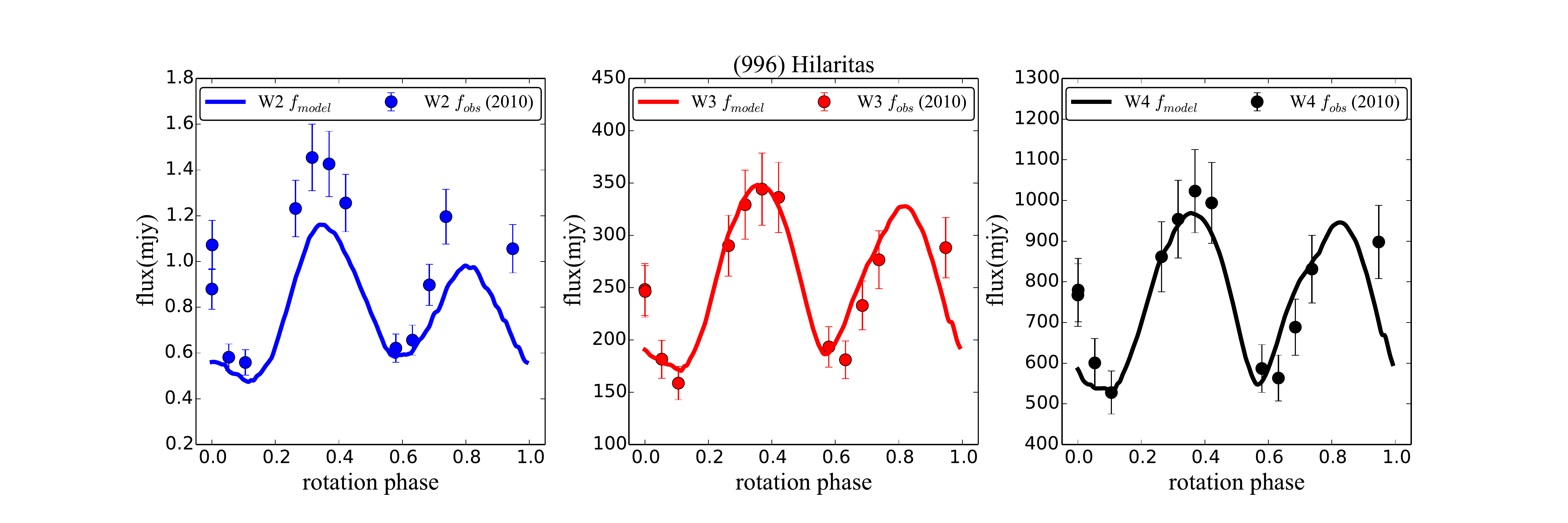}
    \caption{W2, W3 and W4 thermal light curves of (996) Hilaritas.}
    \label{thli996}
\end{figure*}

Using AKARI and WISE observations, \citet{2011PASJ...63.1117U} and \citet{2011ApJ...741...90M} reported its geometric albedo to be $0.069\pm0.008$ and $0.0824\pm0.0180$, respectively, producing an effective diameter of $\sim 33.67\pm1.8~\rm km$ and $30.902\pm0.417~\rm km$. Using 138 WISE/NEOWISE observations (112 in W2, 13 in W3 and 13 in W4), we present a smaller effective diameter of (996) Hilaritas to be $27.560_{-1.925}^{+1.767}~\rm km$ as compared with those of the former studies, thus with a geometric albedo $0.0770_{-0.0090}^{+0.0120}$. From Figure~\ref{chi2gaall}(h), we further show a best-fitting value of $\Gamma = 50_{-26}^{+16}$~$\rm J m^{-2} s^{-1/2} K^{-1}$ and  $f_{\rm r}=0.5_{-0.3}^{+0.0}$. Figure~\ref{thli996} displays the calculated thermal light curves versus observations at W2, W3 and W4 for (996) Hilaritas, revealing that the model are in good accordance with the data at W3 and W4, but is somewhat underestimated at W2. Here we obtain $\chi^2_{\rm min}=8.555$.

\subsection{(1082) Pirola}
\begin{figure*}
  \centering
  \includegraphics[scale=0.40]{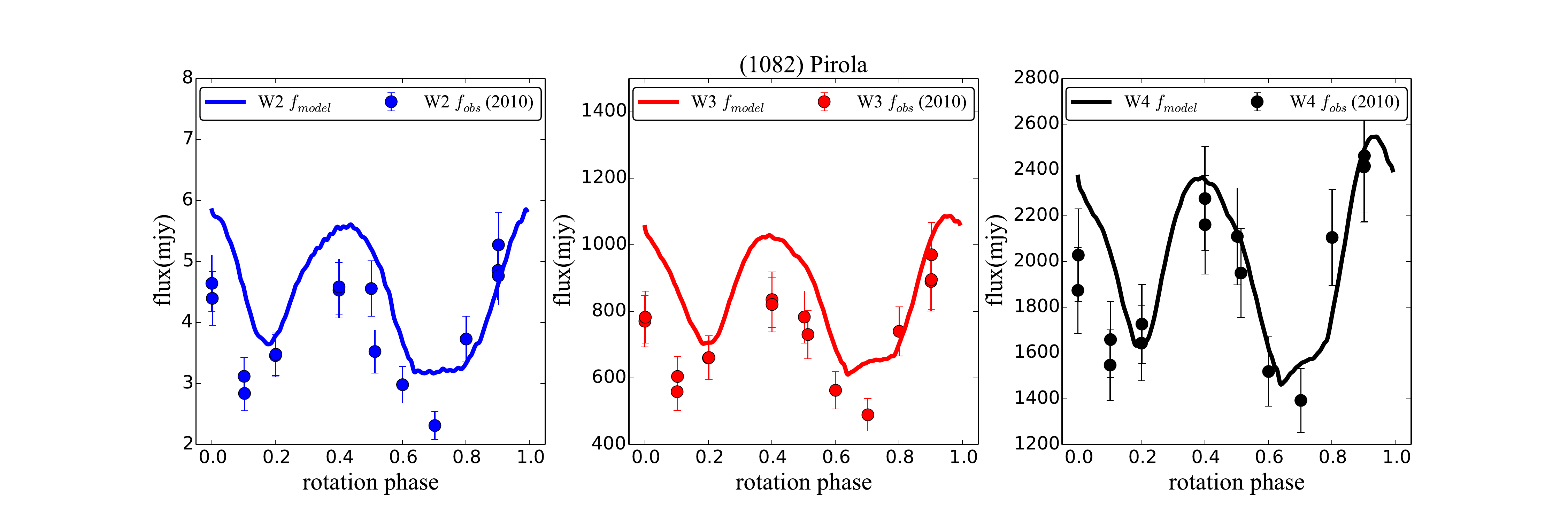}
    \caption{W2, W3 and W4 thermal light curves of (1082) Pirola.}
    \label{thli1082}
\end{figure*}

The previous exploration showed that the geometric albedo of (1082) Pirola varies from $0.052$ to $0.867$, with an effective diameter in the range $37.363\sim48.378~\rm km$ \citep{2011ApJ...741...90M,2011PASJ...63.1117U,2016PDSS..247.....M,2015ApJ...814..117N}. In this work, 189 WISE/NEOWISE observations (157 in W2, 16 in W3 and 16 in W4) are adopted to calculate its thermal parameters. As shown in Figure~\ref{chi2gaall}(i), the best-fitting value of $\Gamma$ and $f_{\rm r}$ is given to be $45_{-12}^{+14}$~$\rm J m^{-2} s^{-1/2} K^{-1}$ and  $f_{\rm r}=0.5_{-0.1}^{+0.0}$. The $p_{\rm v}$ is evaluated to be $0.0725_{-0.0055}^{+0.0075}$, corresponding to $D_{\rm eff}=41.054_{-1.972}^{+1.651}$. Figure~\ref{thli1082} displays thermal light curves of Pirola  at W2, W3 and W4 bands, and the modeled fluxes slightly overestimate the W2 and W3 bands data, leading to a minimum $\chi^2$ of 5.621.

\subsection{(1576) Fabiola}
\begin{figure*}
  \centering
  \includegraphics[scale=0.40]{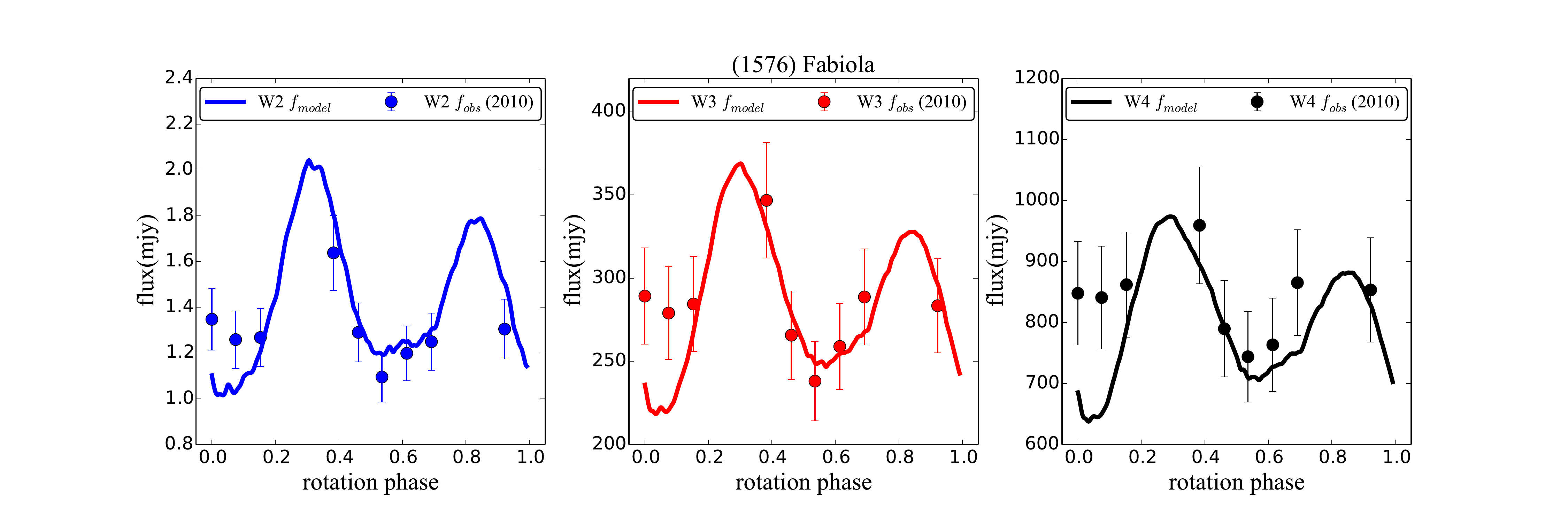}
    \caption{W2, W3 and W4 thermal light curves of (1576) Fabiola.}
    \label{thli1576}
\end{figure*}

On the basis of IRAS, AKARI and WISE/NEOWISE measurements, the geometric albedo of (1576) Fabiola were given to be $0.0746\sim0.123$, and the effective diameter $21.33\sim30.150~\rm km$~\citep{2004PDSS...12.....T,2011PASJ...63.1117U,2011ApJ...741...90M, 2016A&A...591A..14A}. Here we use the ATPM model in combination with WISE/NEOWISE observations (92 in W2, 9 in W3 and 9 in W4) to derive its thermal parameters, and we present the best-fitting  $\Gamma=10_{-10}^{+27}$~$\rm J m^{-2} s^{-1/2} K^{-1}$,  $f_{\rm r}=0.0_{-0.0}^{+0.3}$ (Figure~\ref{chi2gaall}(j)), $p_{\rm v}=0.0830_{-0.0130}^{+0.0135}$, $D_{\rm eff}= 27.796_{-2.018}^{+2.471}~\rm km$, and a low roughness can better fit the observations. Our results of albedo and diameter for this asteroid are slightly different from those in literature. As seen from Figure~\ref{thli1576}, the thermal light curves can reasonably fit the data at W2-W4 bands, and the $\chi^2_{\rm min}$ value is 3.172.

\subsection{(1633) Chimay}
\begin{figure*}
  \centering
  \includegraphics[scale=0.40]{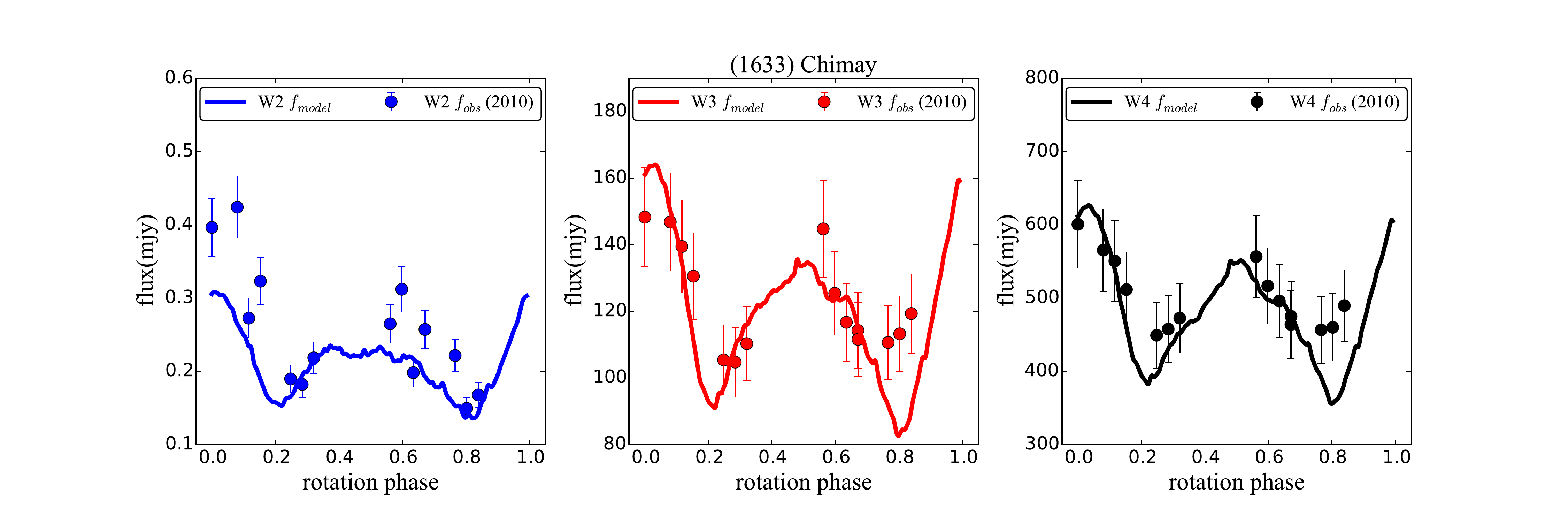}
    \caption{W2, W3 and W4 thermal light curves of (1633) Chimay.}
    \label{thli1633}
\end{figure*}

With AKARI and WISE/NEOWISE observations, (1633) Chimay measures $36.26\pm0.86~\rm km$ and $37.732\pm0.426~\rm km$ in diameter, $0.088\pm0.005$ and $0.0785\pm0.0135$ in geometric albedo~\citep{2011PASJ...63.1117U,2011ApJ...741...90M}. During our fitting, 164 WISE/NEOWISE observations (134 in W2, 15 in W3 and 15 in W4) are utilized to derive its thermal parameters. As shown in Figure ~\ref{chi2gaall}(k), the minimum value of $\chi^2$ is correlated to thermal inertia of $30_{-13}^{+18}$~$\rm J m^{-2} s^{-1/2} K^{-1}$ and roughness fraction of $0.1_{-0.01}^{+0.2}$, with respect to the geometric albedo $p_{\rm v}=0.0405_{-0.0060}^{+0.0065}$ and effective diameter $D_{\rm eff}=47.849_{-3.431}^{+3.993}~\rm km$. Such low value of
the derived geometric albedo is indicative of that (1633) Chimay may be a C-type or B-type asteroid. Here we present the thermal light curves with the data at W2, W3 and W4 bands (Figure~\ref{thli1633}), and the minimum $\chi^2$ is 5.479.

\subsection{(1687) Glarona}
\begin{figure*}
  \centering
  \includegraphics[scale=0.40]{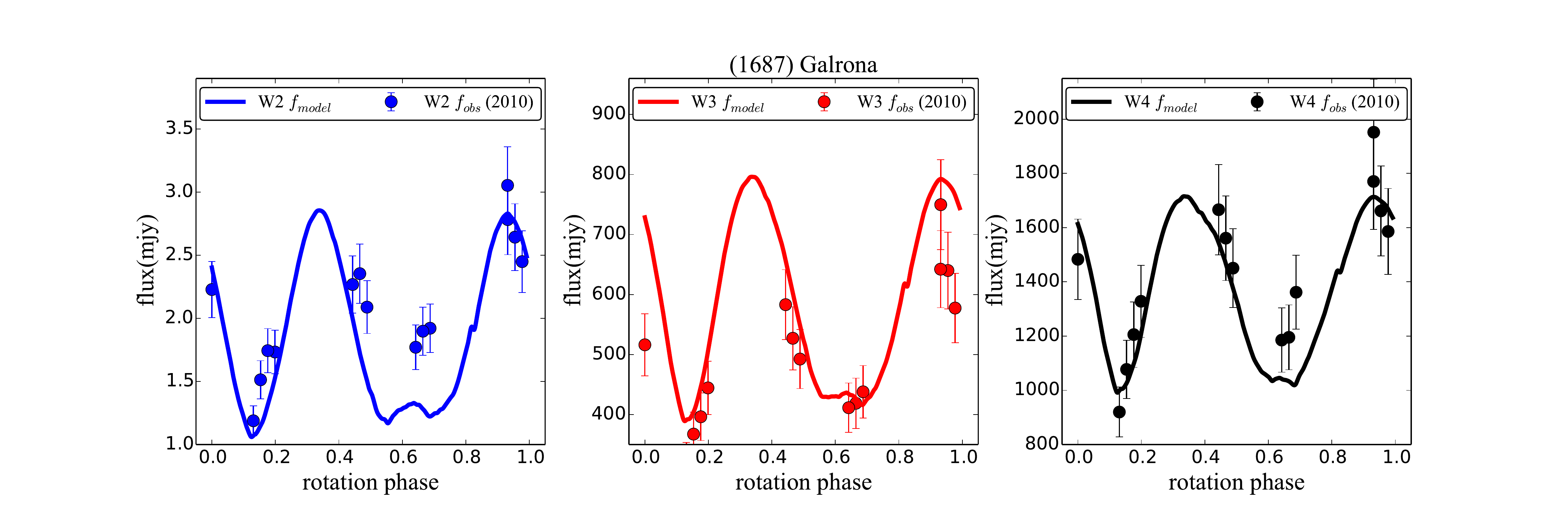}
    \caption{W2, W3 and W4 thermal light curves of (1687) Glarona.}
    \label{thli1687}
\end{figure*}

\citet{2011ApJ...741...90M} presented the albedo and diameter of (1687) Glarona: $p_{\rm v, WISE}=0.0795\pm0.0130$, $D_{\rm eff, WISE}=42.007\pm0.515~\rm km$. Here, 151 WISE observations (121 in W2, 15 in W3 and 15 in W4) are employed in our fitting to understand its thermophysical characteristics. As shown in (Figure~\ref{chi2gaall}(l)), a low thermal inertia of $\Gamma=15_{-10}^{+23}$~$\rm J m^{-2} s^{-1/2} K^{-1}$, as well as a low roughness of $0.0_{-0.0}^{+0.5}$ can be given, with respect to a $\chi^2_{\rm min}=6.475$. The geometric albedo is derived to be $p_{\rm v}=0.1155_{-0.0090}^{+0.0065}$, with an effective diameter $D_{\rm eff}=34.852_{-0.941}^{+1.442}~\rm km$. Our results of $p_{\rm v}$ and $D_{\rm eff}$ slightly vary from those of \citet{2011ApJ...741...90M} in that it may be induced by the usage of ATPM. The W2, W3 and W4 bands thermal light curves are exhibited in Figure~\ref{thli1687}, with the $\chi^2_{\rm min}$ of 6.475.

\subsection{(1691) Oort}
\begin{figure*}
  \centering
  \includegraphics[scale=0.40]{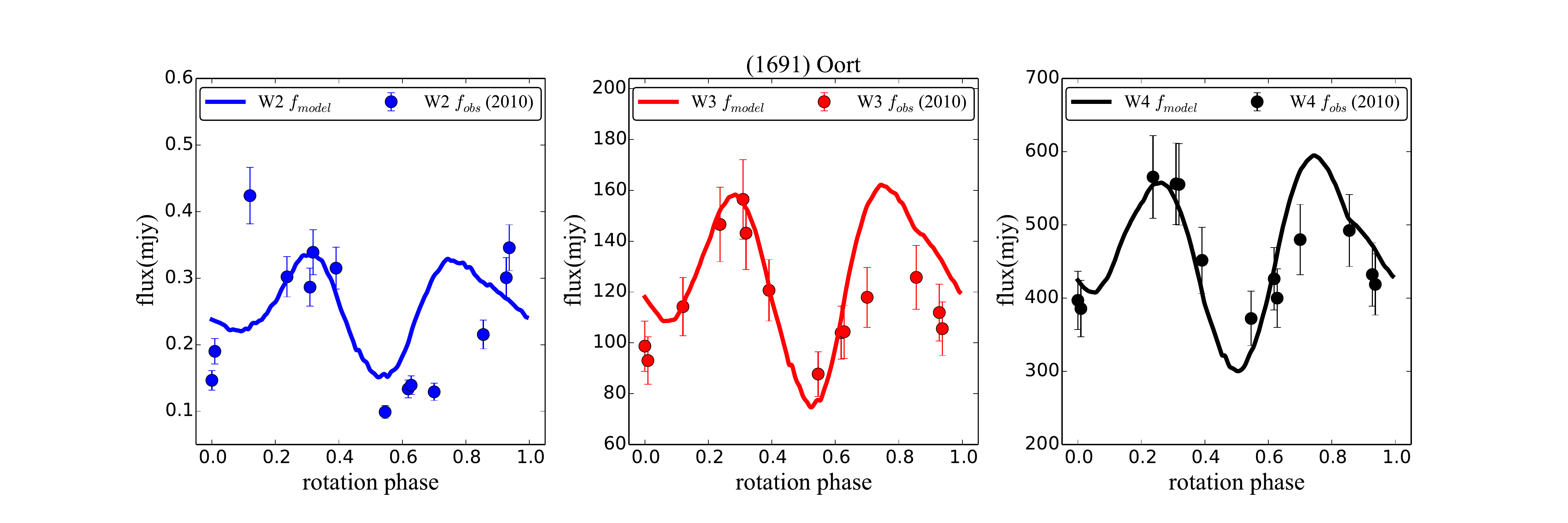}
    \caption{W2, W3 and W4 thermal light curves of (1691) Oort.}
    \label{thli1691}
\end{figure*}

The previous studies showed that Oort's effective diameter was measured to be $33.6\sim37.37 \rm~km$, and geometric albedo ranges from 0.053 to 0.065~\citep{2011ApJ...741...90M,2011PASJ...63.1117U,2014ApJ...791..121M}. In this work, we employ 157 WISE/NEOWISE observations (129 in W2, 14 in W3 and 14 in W4) to calculate the thermal parameters of the asteroid, and find that a low thermal inertia is confined to be $10_{-10}^{+19}$~$\rm J m^{-2} s^{-1/2} K^{-1}$ with a medium roughness fraction of $0.3_{-0.1}^{+0.2}$~(Figure~\ref{chi2gaall}(m)). The value of $\Gamma$ is close to that of \citet{2018Icar..309..297H}, of $22\pm22$~$\rm J m^{-2} s^{-1/2} K^{-1}$. Moreover, we further evaluate the geometric albedo of $0.0635_{-0.0085}^{+0.0085}$ and the effective diameter of $34.884_{-2.121}^{+2.595}~\rm km$. The thermal light curves are shown in Figure~\ref{thli1691}. Our ATPM results provide acceptable fits at W4 band, but seem to slightly deviate from the observations at W2 and W3, with $\chi^2_{\rm min}=7.9214$.

\subsection{(2528) Mohler}
\begin{figure*}
  \centering
  \includegraphics[scale=0.40]{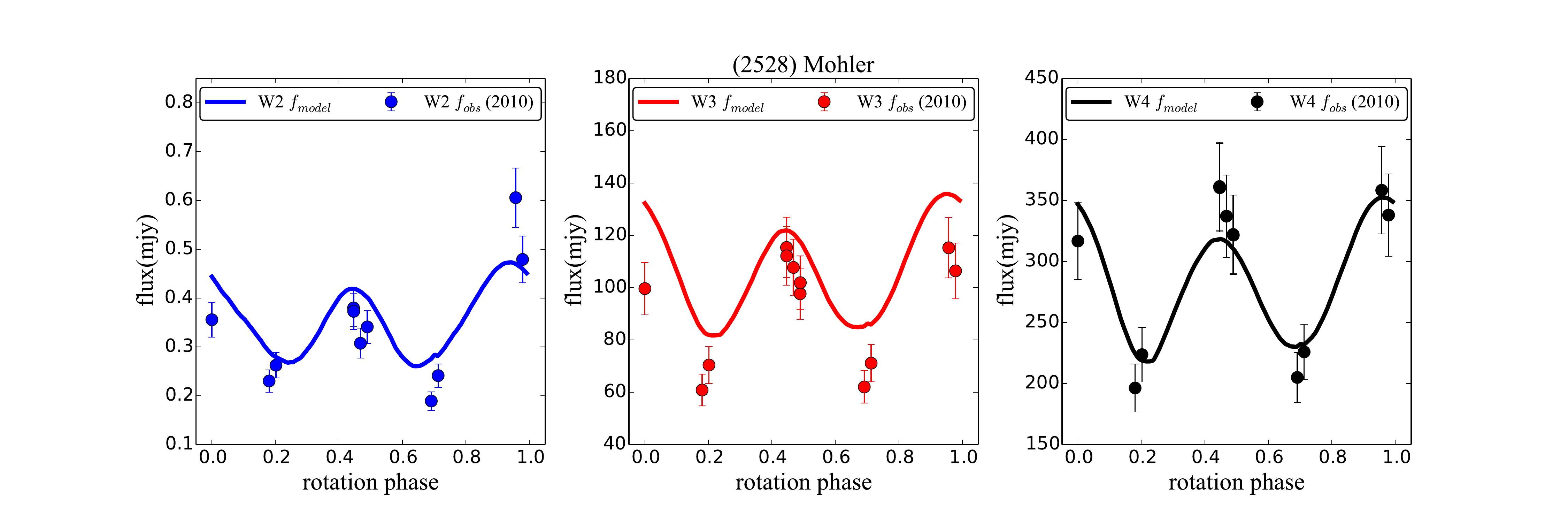}
    \caption{W2, W3 and W4 thermal light curves of (2528) Mohler.}
    \label{thli2528}
\end{figure*}

There are comparatively fewer WISE/NEOWISE data points that are available for this object (45 in W2, 12 in W3, 12 in W4). According to Figure~\ref{chi2gaall}(n), the minimum $\chi^2$ of 10.0837 indicates a relatively poor fit, and the corresponding thermal inertia is $35_{-28}^{+15}$~$\rm J m^{-2} s^{-1/2} K^{-1}$, with respect to a relatively high roughness of $0.5_{-0.5}^{+0.0}$. However, our results of albedo and diameter are consistent with the typical values, $p_{\rm v}=0.0764_{-0.0090}^{+0.0063}$, $D_{\rm eff}=16.826_{-0.654}^{+1.088}~\rm km$, which are similar to those of \citet{2011ApJ...741...90M}, who obtained $p_{\rm v} = 0.0567\pm0.0045$ and $D_{\rm eff} = 19.443\pm0.121~\rm km$. Figure~\ref{thli2528} shows the thermal light curves plotted against the observations at W2, W3 and W4, although the theoretical fluxes can conduct good fitting with the data at W4 band, the fit is overestimated at W2 and W3 with a $\chi^2_{\rm min}$ of 10.084, which indicates a relatively poor fitting degree.

\subsection{(2592) Hunan}
\begin{figure*}
  \centering
  \includegraphics[scale=0.40]{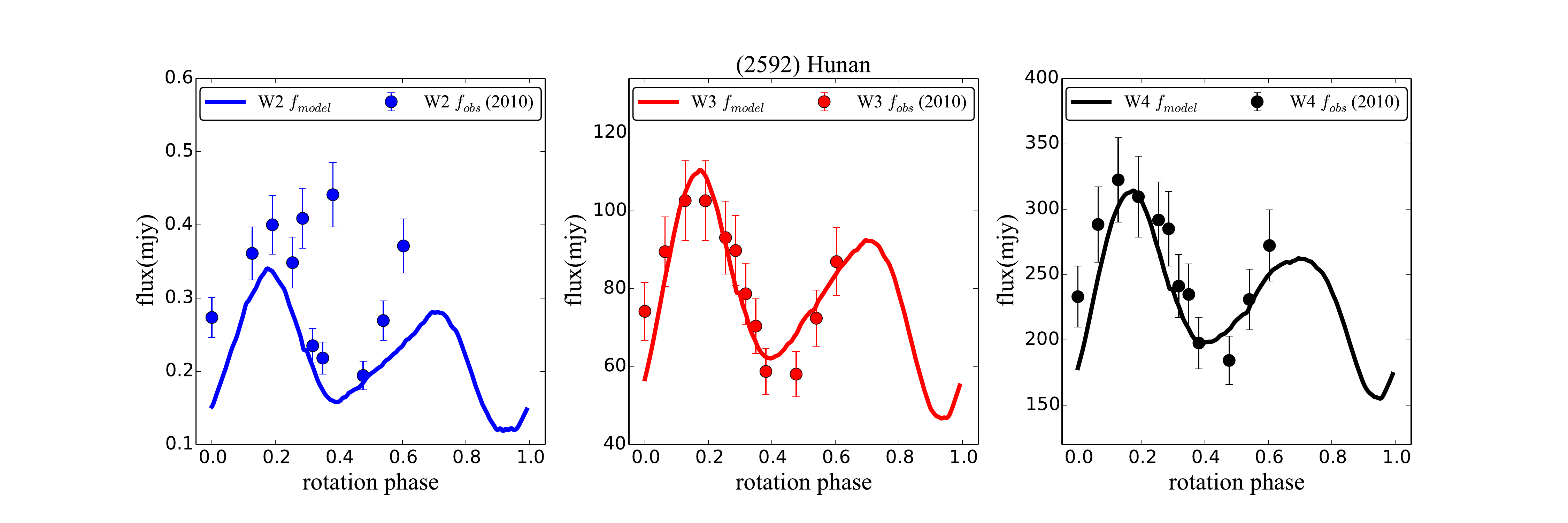}
    \caption{W2, W3 and W4 thermal light curves of (2592) Hunan.}
    \label{thli2592}
\end{figure*}

The WISE/NEOWISE observations combined with the NEATM model give the albedo and diameter of Hunan ranges from $0.072\sim0.08$ and $15.260\sim18.533~\rm km$ \citep{2011ApJ...741...90M,2016AJ....152...63N,2016PDSS..247.....M}. In our thermal modeling process, we adopt 57 W2, W3 and W4 WISE/NEOWISE observations (33 in W2, 12 in W3 and 12 in W4). The best-fitting values are $\Gamma=40_{-40}^{+46}$~$\rm J m^{-2} s^{-1/2} K^{-1}$, $f_{\rm r}=0.3_{-0.3}^{+0.2}$, $p_{\rm v}=0.0636_{-0.0060}^{+0.0087}$ and $D_{\rm eff}=18.958_{-1.177}^{+0.963}~\rm km$. The results of $p_{\rm v}$ and $D_{\rm eff}$ are similar to those of \citet{2016PDSS..247.....M}. We plot the W2, W3 and W4 thermal light curves in Figure~\ref{thli2592}. As can be seen from Figure~\ref{thli2592}, the calculated ATPM fluxes perform a reasonable fitting at W3 and W4 bands, but underestimate W2 observations, where the minimum $\chi^2$ is 6.823.

\subsection{(2659) Millis}
\begin{figure*}
  \centering
  \includegraphics[scale=0.40]{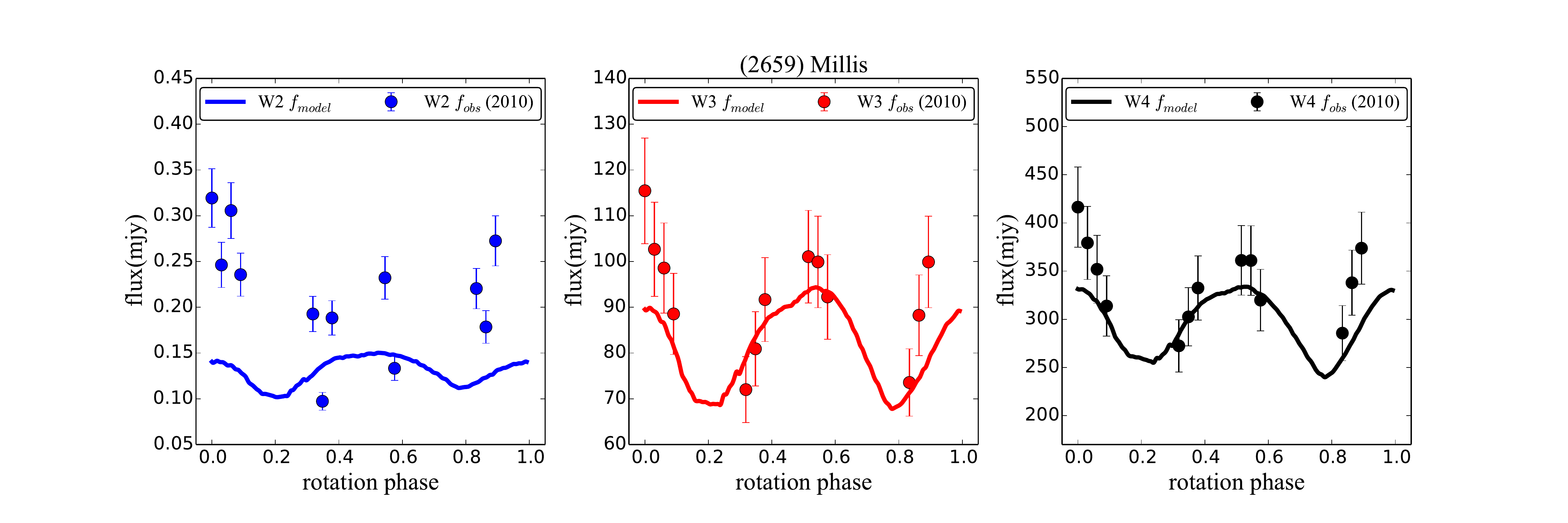}
    \caption{W2, W3 and W4 thermal light curves of (2659) Millis.}
    \label{thli2659}
\end{figure*}

The albedo and diameter of Millis from NEATM spans from $0.0498$ to $0.071$ and $26.42\sim29.53~\rm km$, respectively \citep{2004PDSS...12.....T,2011PASJ...63.1117U,2016A&A...591A..14A,2011ApJ...741...90M,2015ApJ...814..117N}. Here, we use 140 WISE/NEOWISE observations (114 in W2, 13 in W3 and 13 in W4) to investigate the thermal parameters for (2659) Millis. The $\Gamma-\chi^2$ profile is plotted in Figure~\ref{chi2gaall}(p),  where a $\chi^2_{\rm min}=6.2311$ is relevant to a best-fitting thermal inertia  $\Gamma=35_{-16}^{+19}$~$\rm J m^{-2} s^{-1/2} K^{-1}$, a roughness fraction $0.5_{-0.3}^{+0.0}$, a geometric albedo $p_{\rm v}=0.0645_{-0.0072}^{+0.0125}$, and an effective diameter $D_{\rm eff}=27.463_{-2.328}^{+1.674}~\rm km$. Figure~\ref{thli2659} exhibits that the thermal light curves versus the data at W2, W3 and W4, however, the ATPM fluxes at W2 deviate a lot from the observations, but are roughly consistent with those at W3 and W4 bands.

\subsection{(2673) Lossignol}
\begin{figure*}
  \centering
  \includegraphics[scale=0.40]{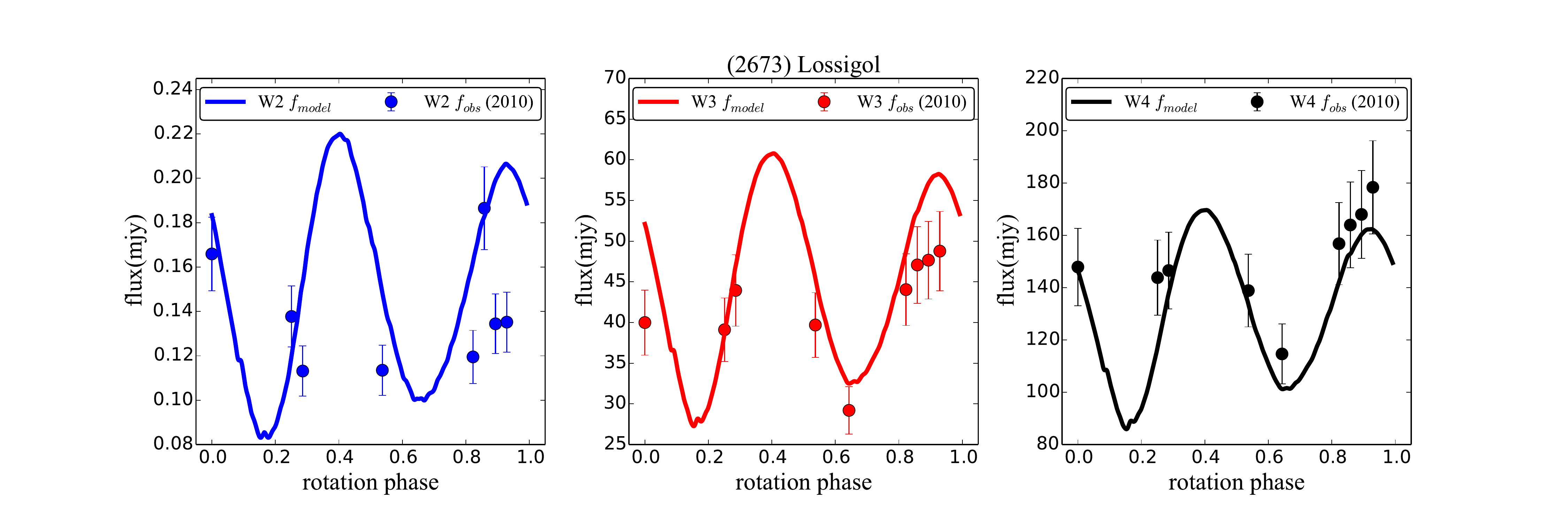}
    \caption{W2, W3 and W4 thermal light curves of (2673) Lossignol.}
    \label{thli2673}
\end{figure*}

With WISE/NEOWISE observations and NEATM model, the albedo and diameter of (2673) Lossignol were determined to be 0.077 and 15.119~km \citep{2011ApJ...741...90M,2016PDSS..247.....M}. In our modeling, there are simply 29 WISE/NEOWISE data points are adopted (11 in W2, 9 in W3, 9 in W4). From Figure~\ref{chi2gaall}(q), we obtain a best-fitting solution of $\Gamma=15_{-15}^{+29}$~$\rm J m^{-2} s^{-1/2} K^{-1}$, and $f_{\rm r}=0.2_{-0.2}^{+0.3}$. The geometric albedo and effective diameter are estimated to be $0.0860_{-0.0147}^{+0.0117}$ and $14.005_{-0.865}^{+1.376}~\rm km$, respectively. The outcomes of $p_{\rm v}$ and $D_{\rm eff}$ are in agreement with the literature results. Figure~\ref{thli2673} shows 3-bands thermal light curves, and the value of $\chi^2_{\rm min}$ is 5.967.

\subsection{(2708) Burns}
\begin{figure*}
  \centering
  \includegraphics[scale=0.40]{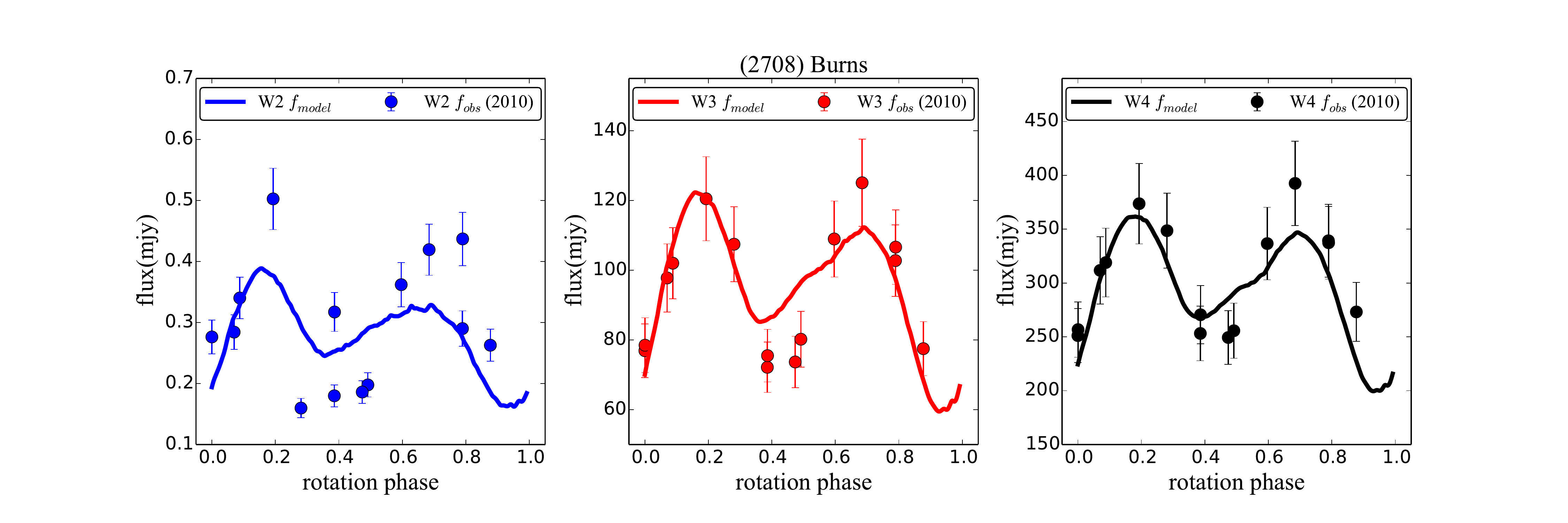}
    \caption{W2, W3 and W4 thermal light curves of (2708) Burns.}
    \label{thli2708}
\end{figure*}

Asteroid (2708) Burns is a B-type Themistian with literature geometric albedo that measures from $0.06\sim0.12$ and diameter in the range of $13.63\sim22~\rm km$ \citep{2013A&A...554A..71A,2015ApJ...814..117N,2016A&A...591A..14A,2016PDSS..247.....M,2011ApJ...741...90M}. Here we totally use 93 WISE/NEOWISE observations (63 in W2, 15 in W3 and 15 in W4) to evaluate its thermal parameters. From the $\chi^2-\Gamma$ profile in Figure~\ref{chi2gaall}(r), we constrain the thermal inertia of (2708) Burns to be $65_{-48}^{+25}$~$\rm J m^{-2} s^{-1/2} K^{-1}$, as well as a relatively high roughness of $0.5_{-0.5}^{+0.0}$. In addition, the geometric albedo and effective diameter are estimated to be $p_{\rm v}=0.0570_{-0.0097}^{+0.0110}$, $D_{\rm eff}=20.492_{-1.731}^{+2.003}~\rm km$. The thermal light curves are plotted in Figure~\ref{thli2708}, we obtain a relatively large $\chi^2_{\rm min}$ value of 8.2179, because the W2 thermal light curve performs a poor fit.

\subsection{(2718) Handley}
\begin{figure*}
  \centering
  \includegraphics[scale=0.40]{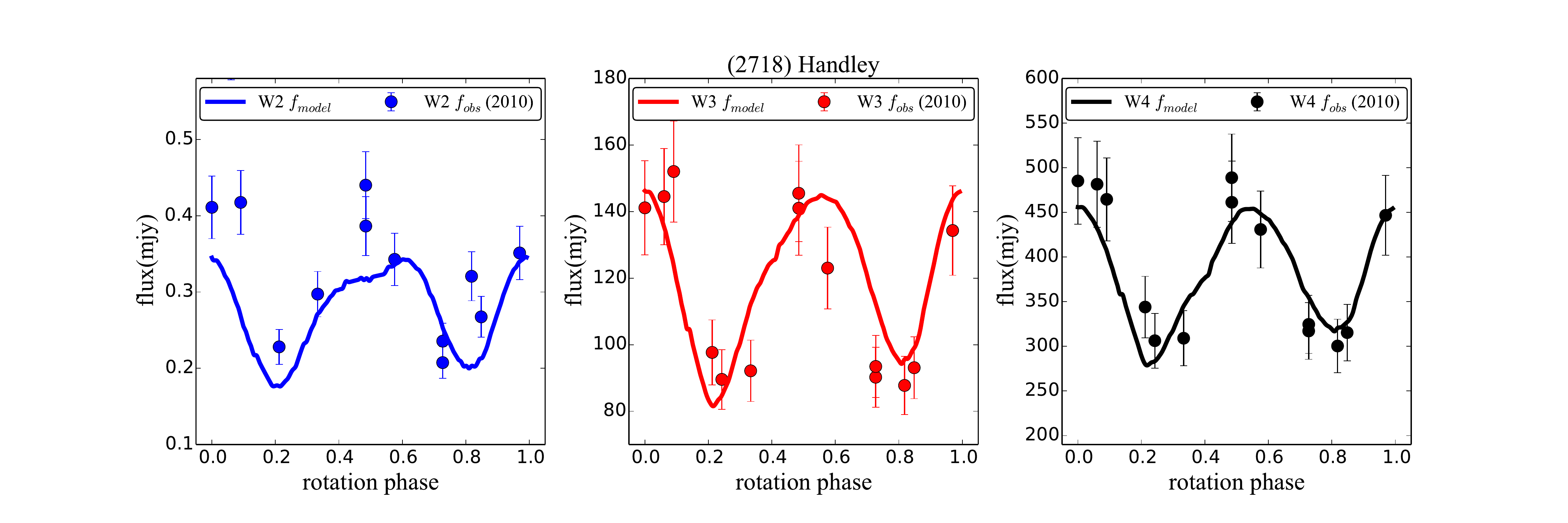}
    \caption{W2, W3 and W4 thermal light curves of (2718) Handley.}
    \label{thli2718}
\end{figure*}

The AKARI and WISE/NEOWISE observations measure similar geometric albedo of Handley to be $0.055\sim0.058$, and the diameter ranges from $25.309\sim25.929~\rm km$ \citep{2011PASJ...63.1117U,2011ApJ...741...90M,2016PDSS..247.....M}. In our fitting, 114  WISE/NEOWISE measurements are adopted (86 in W2, 14 in W3, 14 in W4) to further investigate its thermal nature. From Figure~\ref{chi2gaall}(s), we can see that a minimum value of $\chi^2$ is correlated to thermal inertia of $30_{-30}^{+16}$~$\rm J m^{-2} s^{-1/2} K^{-1}$ and roughness of  $0.5_{-0.4}^{+0.0}$. The geometric albedo is constrained to be $0.0519_{-0.0073}^{+0.0038}$, and an effective diameter  $24.431_{-0.849}^{+1.924}~\rm km$. Moreover, for this asteroid,  Figure~\ref{thli2718} indicates that our computed fluxes perform a close matching with the infrared data, and we have derived the minimum $\chi^2$ of 8.069.

\subsection{(2803) Vilho}
\begin{figure*}
  \centering
  \includegraphics[scale=0.40]{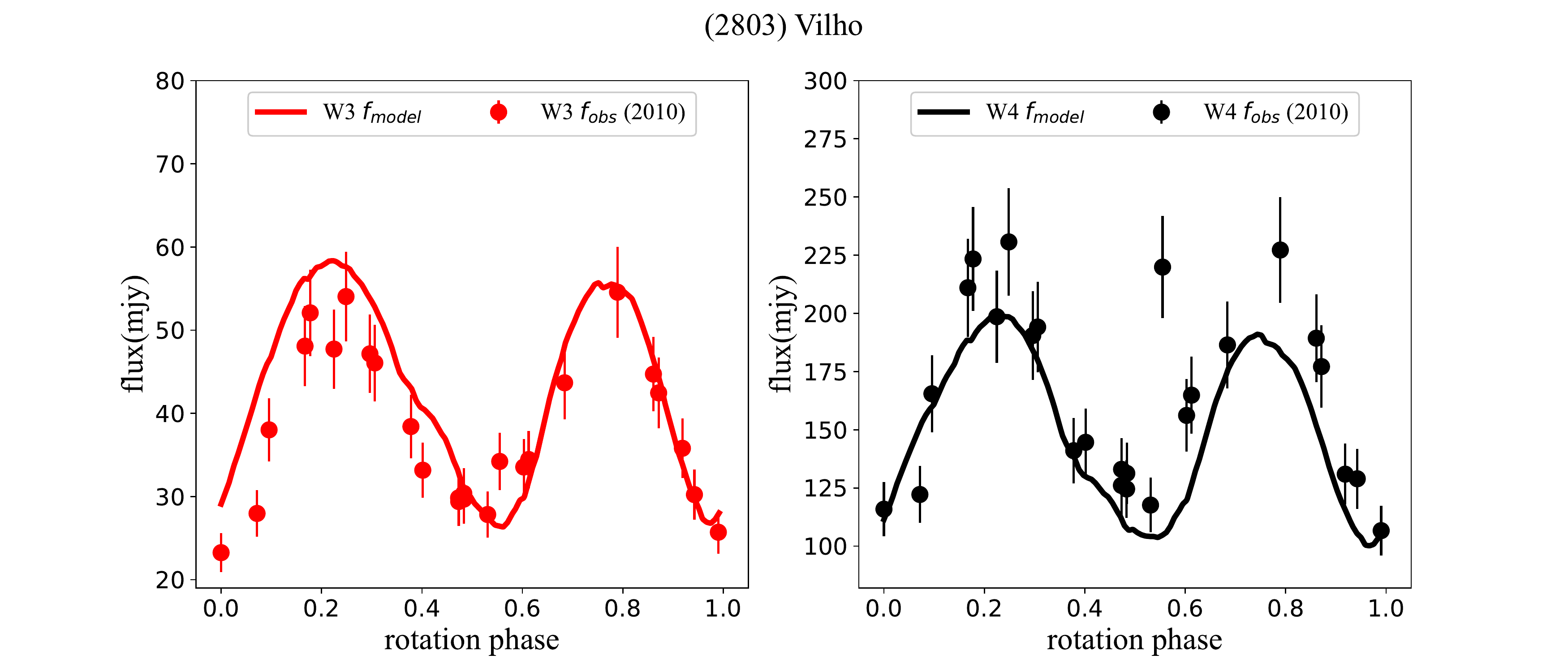}
    \caption{w3 and W4 thermal light curves of (2803) Vilho.}
    \label{thli2803}
\end{figure*}

For C-type Themistian (2803) Vilho, the literature results of albedo and diameter ranges from $0.068\sim 0.1$ and $17.72\sim22.96~\rm km$ \citep{2011PASJ...63.1117U,2011ApJ...741...90M, 2015ApJ...814..117N,2016PDSS..247.....M}. In this work, we utilize 52 WISE/NEOWISE observations (26 in W3 and 26 in W4) for fitting. The best-fitting solution gives a comparatively high thermal inertia of $110_{-29}^{+12}$~$\rm J m^{-2} s^{-1/2} K^{-1}$ and high roughness of  $0.5_{-0.2}^{+0.0}$ (see Figure~\ref{chi2gaall}(t)) with a minimum $\chi^2$ value 5.217. The geometric albedo is derived to be $0.0360_{-0.0010}^{+0.0071}$, and $D_{\rm eff}=27.757_{-2.389}^{+0.394}~\rm km$. The derived geometric albedo and effective diameter of (2803) Vilho are quite different from those of previous works. We show that (2803) Vilho is the only asteroid that bears thermal inertia larger than 100 ~$\rm J m^{-2} s^{-1/2} K^{-1}$. Figure~\ref{thli2803} displays the thermal light curves at W3 and W4 bands, where the ATPM fluxes coincide with the observations, and the $\chi^2_{\rm min}$ is calculated to be 5.218.

%%OK

\section{Thermal parameters}

\subsection{Size Distribution}
\begin{figure*}
  \centering
  \includegraphics[scale=0.38]{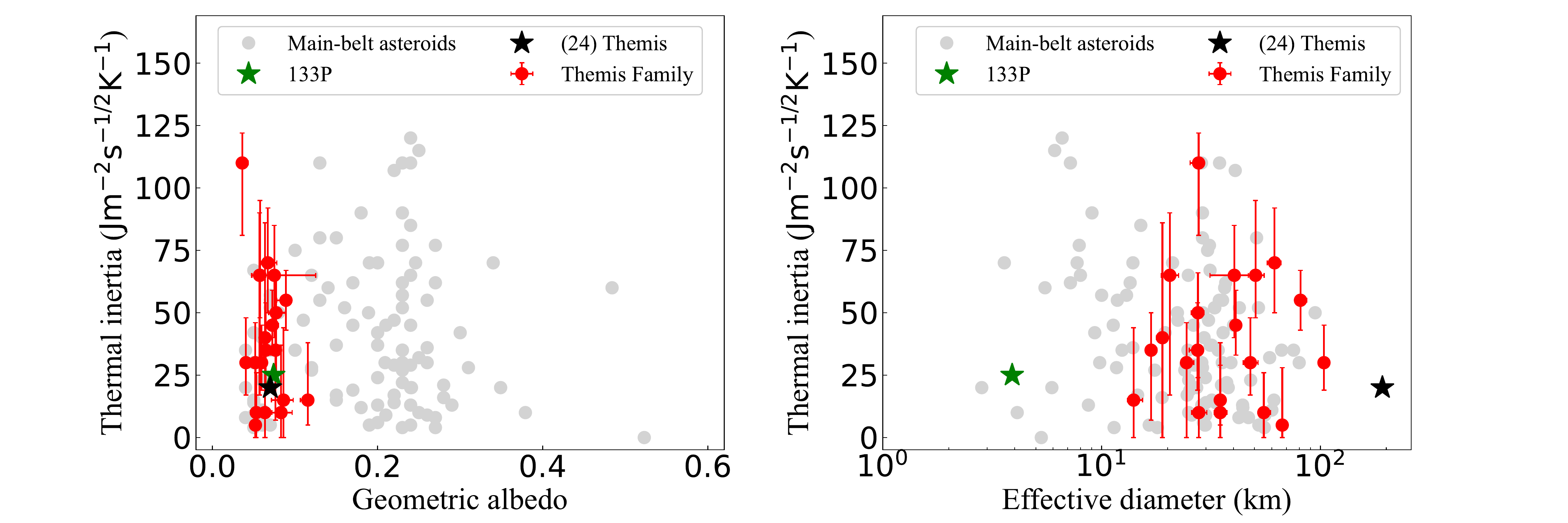}
    \caption{The $\Gamma -p_{\rm v}$ and $\Gamma-D_{\rm eff}$ relationships of Themis family members (\emph{Red dots with error bars}) and the main-belt asteroids (\emph{Gray dots}) in \citet{2018Icar..309..297H}. We also plot the results of (24) Themis (\emph{marked up by black pentagram}) and 133P/Elst-Pizarro (\emph{marked up by green pentagram})  from \citet{2020ApJ...898L..45O,2020AJ....159...66Y}. }
    \label{th_pv_def}
\end{figure*}

With the aid of NEOWISE data, \citet{2013ApJ...770....7M} measured the size-frequency distribution (SFD) for 76 asteroid families. SFD can be expressed by $N\propto D^\alpha$, where $N$ is the number of family members that have diameter larger than $D$, and $\alpha$ is the SFD slope. \citet{2015aste.book..323M} gave the value $\alpha$ of $-2.313\pm0.017$ for Themis family and the SFD range of $7.3\sim55.6~\rm km$. As illustrated in the right panel of Figure~\ref{th_pv_def}, we obtain the range of $D_{\rm eff}$ to be $14.005\sim103.816~\rm km$, where asteroid (171) Ophelia have a maximum effective diameter of $103.816~\rm km$. \citet{2009P&SS...57..259D} provided the power law relationship between thermal inertia and effective diameter to be $\Gamma=d_{\rm 0} D^{-\xi}$, and a best-fitting $d_{0}$ and $\xi$ of $300\pm47$ and $0.48\pm0.04$, respectively, indicating an inversely proportional relationship between $\Gamma$ and diameter. In addition, \citet{2009P&SS...57..259D} derived $\xi$ for MBAs ($1.4\pm0.2$) and NEAs ($0.32\pm0.09$), respectively. However, when the asteroid population is large enough, especially for the main-belt asteroids with a relatively low thermal inertia ($<100~\rm J m^{-2} s^{-1/2} K^{-1}$ ), this inverse relationship becomes unclear (Figure~\ref{th_pv_def}) and requires deeper exploration based on diverse asteroid families.

\subsection{Geometric Albedo}
\begin{figure*}
  \centering
  \includegraphics[scale=0.42]{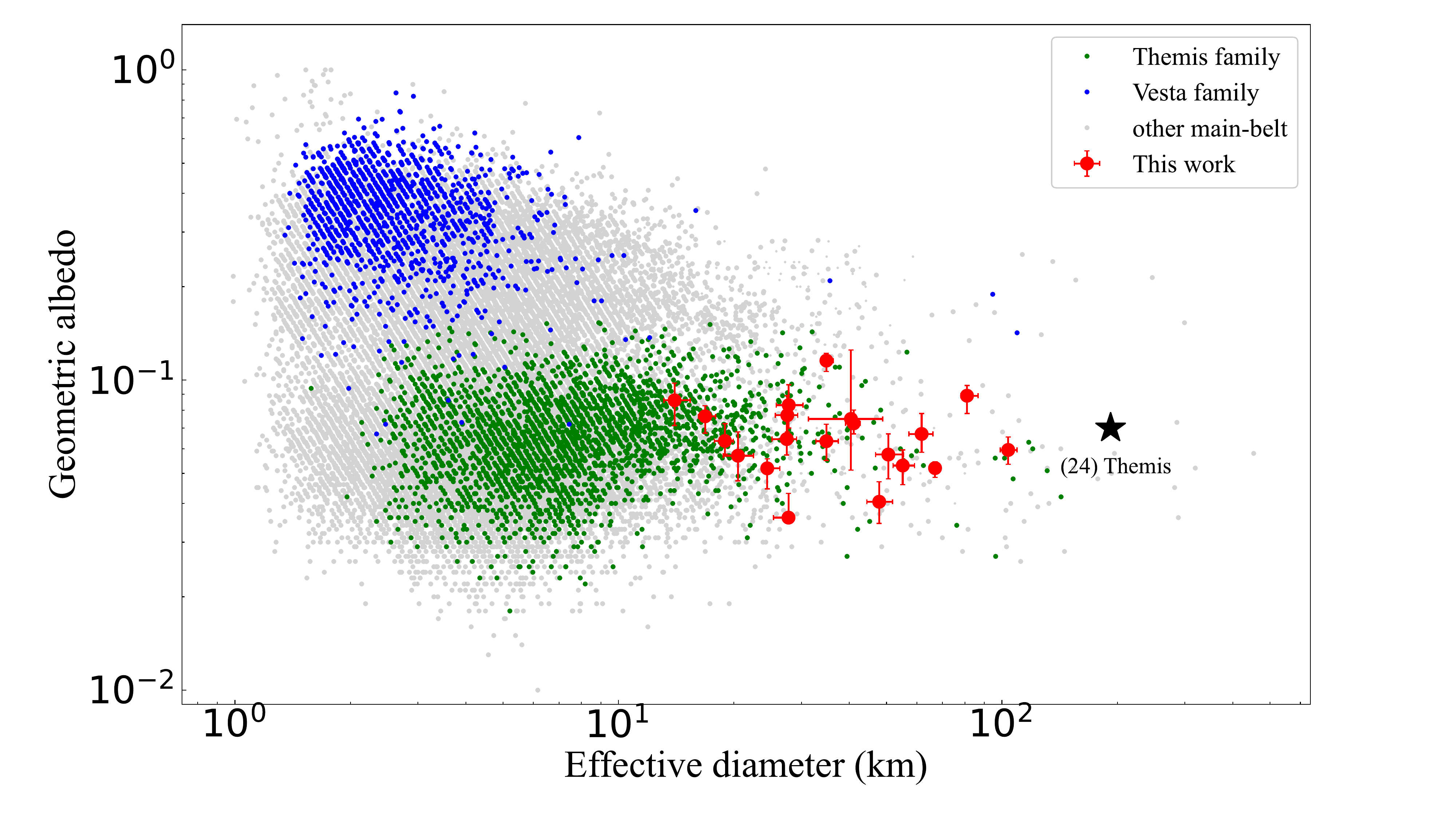}
    \caption{The $p_{\rm v}-D_{\rm eff}$ relationship of main belt asteroids. The Vesta family and Themis family are plotted in blue and green dots, respectively, while other main-belt asteroids are shown by gray dots. Red dots with error bars are the results of this work, where (24) Themis is marked up by black pentagram. Note that each axis is logarithmic, thus the Themis family covers a much smaller range of geometric albedo.}
    \label{pvdeff}
\end{figure*}

The collisional events that formed the asteroid families are expected to produce materials (i.e., the family members) from the parent body. For a heterogeneous parent body, a wide range of color indexes, albedos and spectrals are expected, while for a homogeneous parent, the resultant family members usually have relatively narrow range of these physical parameters \citep{2015aste.book..323M}.  \citet{2012A&A...537A..73L} showed that $5-14 ~\rm \mu m$ spectra of 8 Themis family asteroids and obtained a mean albedo of $0.07\pm0.02$ based on NEATM. Furthermore, \citet{2013ApJ...770....7M} gave a mean value of $p_{\rm v}=$ $0.066\pm0.021$ for Themis family members.  In this work, we obtain an average geometric albedo of Themis family to be $0.067\pm0.018$, which is in good agreement with that of \citet{2013ApJ...770....7M}. Figure~\ref{pvdeff} exhibits the distribution of $p_{\rm v}$ and $D_{\rm eff}$, where red error bars indicate the results of our work, which are similar to the $p_{\rm v}$ of their parent body (24) Themis. We further superimpose the MBAs' albedos from \citet{2013ApJ...770....7M,2018Icar..309..297H,jianghx2019} (marked up by light gray dots), and we offer other Themis family asteroids in \citet{2013ApJ...770....7M} by green dots. Most of Themistians appear to have relatively low geometric albedos, which agrees with the typical values of C-type asteroids. As a comparison, we give the Vesta family's albedo distribution with blue dots in Figure~\ref{pvdeff}, which are retrieved from \citet{2013ApJ...770....7M,jianghx2019}. Unlike the Themis family, the Vesta family's albedo varies in a wide range, implying  a heterogeneous parent body and differentiated surface layers through their long-term evolution.

%Moreover, the carbonaceous asteroids (C-type, which have very low albedos) dominate the outer region of the asteroid belt, while S-type asteroids (silicate-rich) that have relatively higher albedos usually locate at the inner region of the main-belt. Thus, Figure~\ref{semipv} shows a decreasing trend of $p_{\rm v}$ with an increasing semi-major axis.

% The Hungaria family is mainly composed of E-type members that have extremely enstatite surfaces and high albedos.

\begin{figure*}
  \centering
  \includegraphics[scale=0.42]{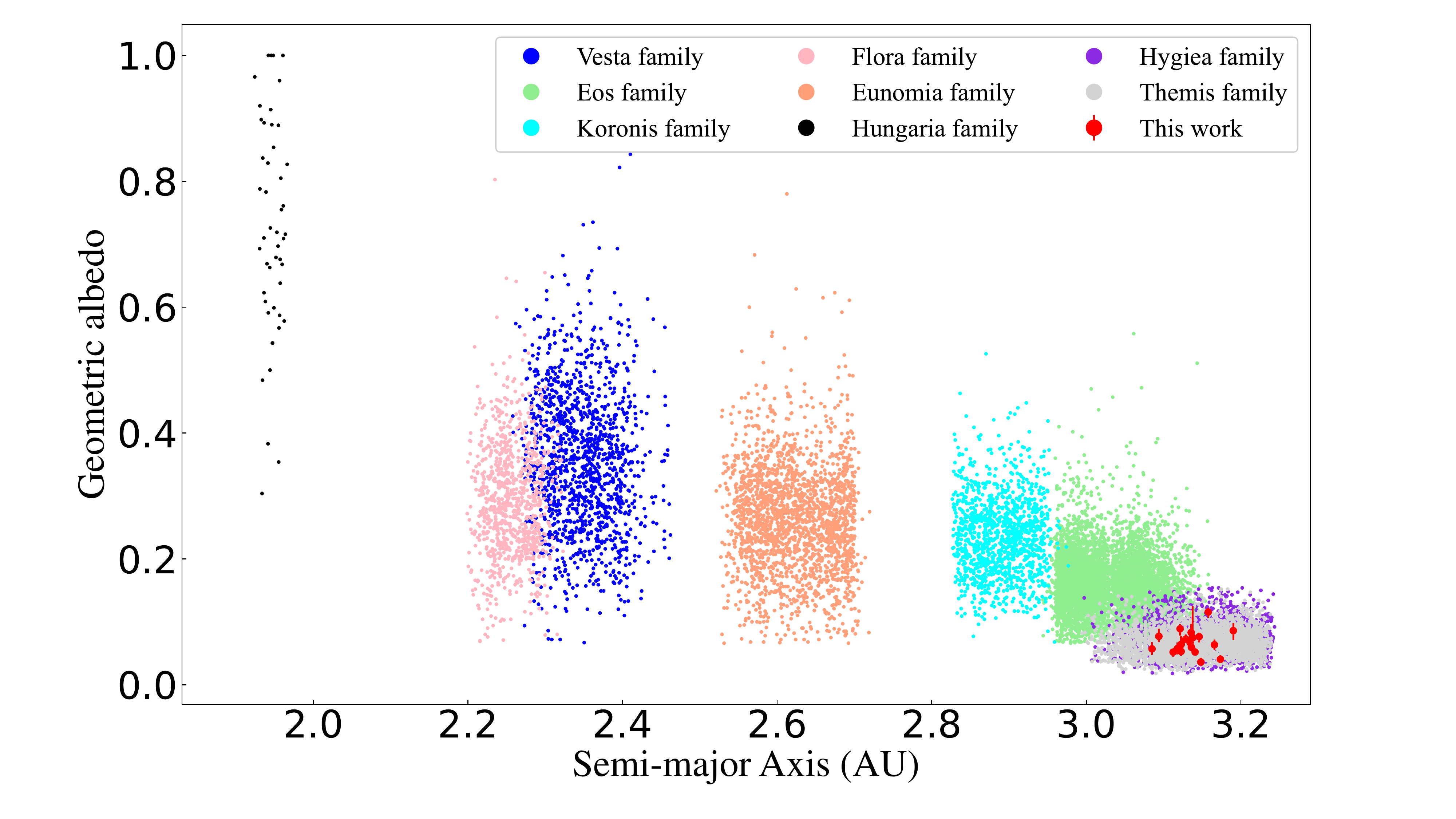}
    \caption{Geometric albedo versus semi-major axis (in AU) for the Themis family members and several prominent families in \citet{2013ApJ...770....7M,2020AJ....159..264J}. The Vesta family and Hungaria family have a wide range of geometric albedo, while for Themis and Hygiea family, the ranges of geometric albedo are relatively smaller.}
    \label{semipv}
\end{figure*}

As a comparison, in Figure~\ref{semipv}, we show our results of geometric albedos against their semi-major axis (red error bars) as well as those of other prominent asteroid families in \citet{2013ApJ...770....7M}.  As shown in Figure~\ref{semipv}, the families like Themis or Hygiea are mainly composed of B- or C-type members, which are located in the outer asteroid belt, and simply cover a very small range of $p_{\rm v}$ compared to other asteroid families (here we obtain a minimum and maximum value of $p_{\rm v}$ to be 0.0360 and 0.1155 for the Themistians under study, respectively). This is consistent with the inference that carbonaceous (C-type, B-type, etc, which have low albedos) asteroids dominate the outer region of the asteroid belt \citep{2007AJ....133.1609W}.  To our best knowledge, C-type asteroids are believed to have primordial components, considering the fact that asteroids usually have undergone considerable evolution processes since their formation, such as space weathering, surface morphology, etc. Here we may infer that the Themis family, which is mainly composed of C-type or B-type asteroids, seems to be a relatively antique family that have not experienced significant migration arising from gravitational perturbation of giant planets after they are separated from their parent body Themis due to collision events at early stage.

\subsection{Thermal Inertia}
\begin{figure*}[htbp!]
  \centering
  \includegraphics[scale=0.42]{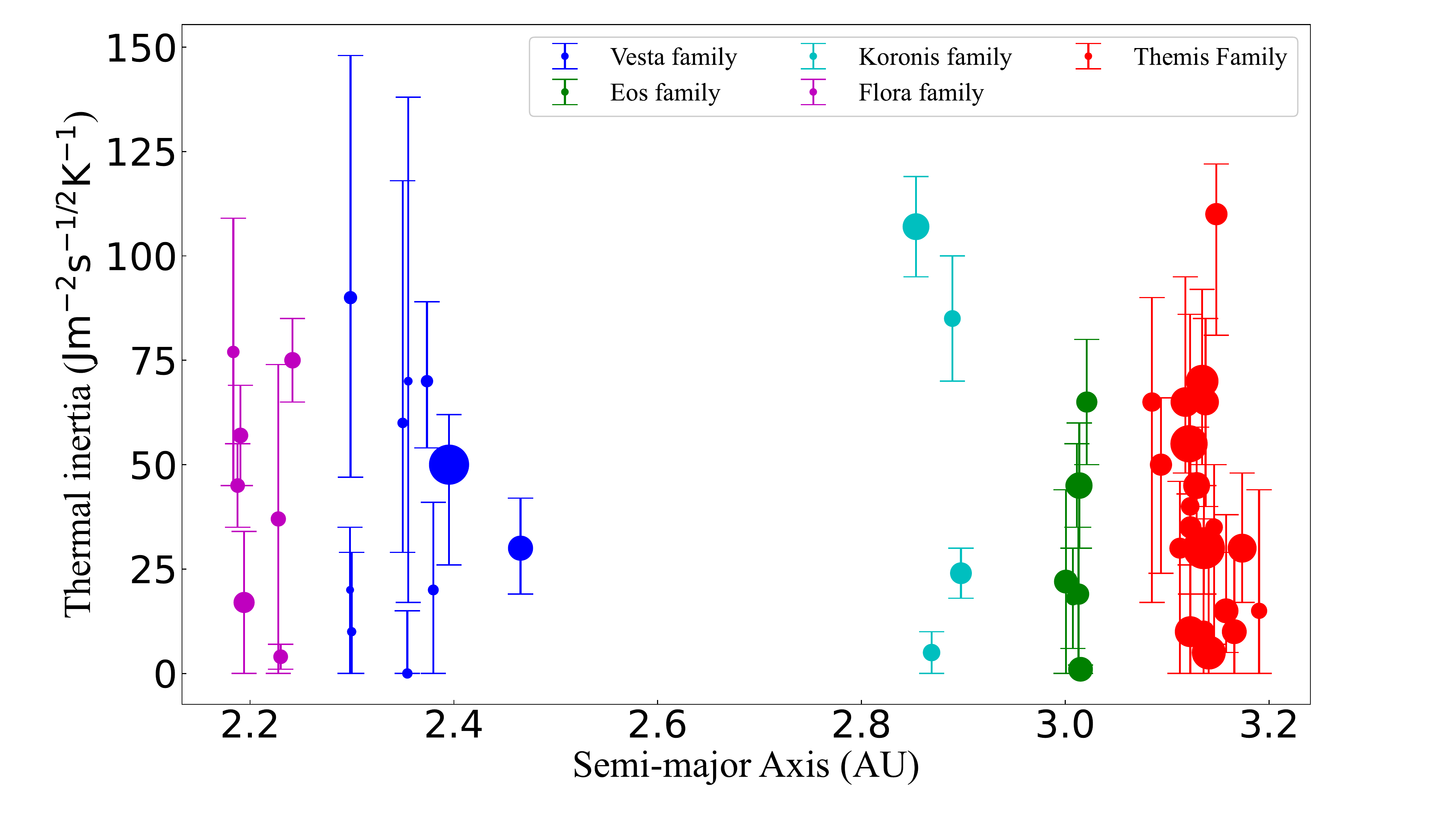}
    \caption{Thermal inertia versus semi-major axis (in AU) for the Themis family in this work and other families in \citet{jianghx2019,2018Icar..309..297H}. The size of points represents the diameter of the asteroid. The red dots with error bars stands for the derived thermal inertia of Themis family.}
    \label{semith}
\end{figure*}

As mentioned above, the Themis family asteroids have roughly primitive materials. Like most of main-belt asteroids, a large number of  Themistians in this work have a very low thermal inertia that are lower than 100~$\rm J m^{-2} s^{-1/2} K^{-1}$, except (2803) Vilho. The average thermal inertia of 20 Themistians is $39.5\pm26.0 \rm J m^{-2} s^{-1/2} K^{-1}$, which is very similar to that of Vesta family of 42~$\rm J m^{-2} s^{-1/2} K^{-1}$ \citep{jianghx2019}, but is a bit larger than that of (24) Themis (marked up by black pentagram in Figure~\ref{th_pv_def}) of \citet{2020ApJ...898L..45O}. Note that (2803) Vilho is the only one that has thermal inertia greater than 100~$\rm J m^{-2} s^{-1/2} K^{-1}$, which obviously differs from others in this study, implying the existence of interlopers in Themis family or this object that may originate from a distinct compositional layer of the parent body. However, the geometric albedo of (2803) Vilho is consistent with the typical value of Themistians, thus we need to take additional clues (such as spectral features) into consideration to determine whether this object can be treated as an 'interloper'.  According to the $\Gamma-p_{\rm v}$ and $\Gamma-D_{\rm eff}$ distribution of main-belt asteroids in Figure~\ref{th_pv_def}, the thermal inertia is relatively evenly distributed between the maximum and minimum values. This phenomenon also occurs in other asteroid families. As shown in Figure~\ref{semith}, red error bars represent the derived thermal inertia $\Gamma$ of this work, whereas those of other families from \citet{2018Icar..309..297H} are plotted with diverse colors, where the size of circles stands for the size of asteroids. However, it is not easy to distinguish different asteroid families by thermal inertia. As described in \citet{2015aste.book..107D}, the asteroid's thermal inertia is associated with the surface temperature, relying on heliocentric distance, thus it can be expressed as \citep{2015aste.book..107D}
\begin{equation}
  \Gamma \propto \sqrt{\kappa} \propto T^{3/2} \propto r^{-3/4}.
  \label{th_helio}
\end{equation}
However, even we normalised the value of thermal inertia into 1 AU from the Sun according to Eq.\ref{th_helio}, the differences in $\Gamma$ distribution between various families are still unclear. This is probably because most main-belt objects have undergone long-term resurface processes. Therefore, although the asteroid families formed at various time, their surfaces may have evolved into similar morphological characteristics (such as fine regolith layers), thereby leading to similar thermal inertia distribution. Subsequently, finely powdered regolith covered on asteroid's surface is a poor heat conductor (as compared with bare rocks or a single particle) because of the existence of tiny intervals between regolith grains thereby inducing a very low thermal inertia. Therefore, according to the value of $\Gamma$, we can infer whether there exist thermally insulating powdered surface materials \citep{2015aste.book..107D}. Furthermore, by using specific thermal conduction model described in \citet{gundlach2013} and the value of thermal inertia, we can further estimate the regolith grain sizes of these Themistians. Considering the volume filling factor of 0.0 to 0.6, and the temperature of 200 K, we obtain the mean regolith grain sizes of these Themistians vary from 0.077 to 8.322 mm, with an average value of $1.616\pm0.494~\rm mm$.

As described above, Themis family is probably closely connected with the active MBAs as well as the main belt comets, e.g., 133P/Elst-Pizarro and 176P/LINEAR. Thus we give the thermal parameters of 133P/Elst-Pizarro \citep{2020AJ....159...66Y} in Figure~\ref{th_pv_def} (marked up by green pentagrams). We find that both thermal inertia and geometric albedo of 133P/Elst-Pizarro are within the range of Themistians we investigated. Using the data from MPC, \citet{2017P&SS..137...52F} reduced 192016 magnitude observations of 165 Themis family asteroids, among which 25 ($15.2\%$) of them exhibit bumps or enhancements in brightness that might suggest low-level cometary activity.  Besides, the activity of asteroid might be triggered by water-ice sublimation, but only a small portion of the Themistians are discovered to have cometary activities. This may be the reason that  different family members originate from different parts of the parent body. Although \citet{2010Natur.464.1320C} predicted that water-ice is widely spread on (24) Themis, several family members may be the fragments that have no water-ice and thus detect no apparent activities. An alternative explanation is that the lifetime of water-ice is much less than the age of the family. As mentioned in \citet{2008ApJ...682..697S}, the existing time of water-ice is strongly affected by temperature of the body (which is mainly concerned with thermal inertia), and the dust/gas production rate is in connection with the effective diameter \citep{2020AJ....159...66Y}. Hence, our results of Themis family members can help us explore their activities in the future work.

\section{Conclusions}
In this work, we apply ATPM combined with  WISE/NEOWISE mid-infrared measurements to investigate the thermal inertia, geometric albedo, effective diameter and roughness fraction of 20 Themis family asteroids. Here we summarize the major results of the asteroids as follows: the average thermal inertia is derived to be $\Gamma_{\rm mean}=$~$39.5\pm26.0 \rm J m^{-2} s^{-1/2} K^{-1}$. The geometric albedo spans from $0.0360$ to $0.1155$, with an averaged value of $p_{\rm v,mean}=0.067\pm0.018$, which agrees well with that of the former study \citep{2013ApJ...770....7M}. The average effective diameter of the investigated Themistians are $41.173\pm22.663~\rm km$. The family members bear a moderate roughness fraction on the surfaces, with a mean value of $0.33\pm0.19$.

Moreover, we present the distribution of these parameters and explore the relation among thermal parameters. The thermal inertia of the Themistians are derived to be relatively small, implying that a fine and mature regolith layer may exist on their surfaces due to long-term space weathering or other effects. In comparison to several prominent families, we find that the $p_{\rm v}$ values of Themistians are rather smaller, and only cover a very small range compared to other prominent families, which is in line with the typical values of B-type and C-type asteroids in main-belt. In addition, for various asteroid families, the value of $p_{\rm v}$ varies notably but may have similar distribution of thermal inertia.  Finally, According to the given diameters of a large portion of main-belt asteroids, the decreasing relationship of $\Gamma-D$ becomes unclear, thereby not following the power law given by \citet{2009P&SS...57..259D}, this is probably due to the small sample population we adopted in this work. Therefore, the similarity in thermal inertia and geometric albedo of Themis members may reveal their close connection in origin and evolution.

However, it should be noteworthy that Themis family are ancient families, which may have been formed 1 Gyr ago, and they are located in the middle or outer region of the main-belt. Thus, it is very important to have a full picture of thermal characteristics for other asteroid populations, e.g., Erigone family that may have much younger age with low geometric albedos and 60\% hydrated, or the Pallas family that is the birthplace of a great many of active asteroids. The forthcoming investigation will enable us to have a better understanding of formation and evolution of the asteroid belt and even the Solar system.

\acknowledgments
We thank two referees for constructive comments and suggestions to improve the manuscript. This work is financially supported by the B-type Strategic Priority Program of the Chinese Academy of Sciences (Grant No. XDB41000000),the National Natural Science Foundation of China (Grant Nos. 12033010, 11661161013, 11633009), CAS Interdisciplinary Innovation Team and Foundation of Minor Planets of the Purple Mountain Observatory. This research has made use of the NASA/IPAC Infrared Science Archive, which is operated by the Jet Propulsion Laboratory, California Institute of Technology, under contract with the National Aeronautics and Space Administration. Research using WISE Release data is eligible for proposals to the NASA ROSES Astrophysics Data Analysis Program.

\bibliography{ms}{}
\bibliographystyle{aasjournal}

\end{document}